\PassOptionsToPackage{table,xcdraw,dvipsnames}{xcolor}
\documentclass{article} 
\usepackage{iclr2025_conference,times}

\usepackage{amsmath,amsfonts,bm}

\def\eqref#1{equation~\ref{#1}}

\def\1{\bm{1}}

\DeclareMathAlphabet{\mathsfit}{\encodingdefault}{\sfdefault}{m}{sl}
\SetMathAlphabet{\mathsfit}{bold}{\encodingdefault}{\sfdefault}{bx}{n}

\usepackage{hyperref}
\usepackage{url}
\usepackage{amsthm}
\usepackage{hyperref}

\usepackage[most]{tcolorbox}
\usepackage{wrapfig}
\usepackage{graphicx,float}
\usepackage{subfigure}
\usepackage{threeparttable}
\usepackage[ruled,vlined]{algorithm2e}
\usepackage{colortbl}
\usepackage{color}
\usepackage{multirow}
\usepackage{tabularx}
\usepackage{float}
\usepackage{graphicx}
\usepackage{booktabs}
\usepackage{arydshln}
\usepackage{enumitem}
\usepackage{wrapfig}
\usepackage{caption}
\usepackage{graphicx}
\usepackage{makecell}
\usepackage{float} 
\usepackage{mathtools}
\usepackage{subfigure}
\usepackage{bbding}
\usepackage{makecell}
\usepackage{tabularx}
\usepackage{amssymb,mathrsfs,amsmath}
\usepackage{pifont}
\usepackage{bbm}

\usepackage{listings}

\definecolor{bittersweet}{rgb}{1.0, 0.44, 0.37}
\definecolor{mygreen}{rgb}{0.29, 0.7, 0.48}
\usepackage{pifont}
\definecolor{demphcolor}{RGB}{144,144,144}

\definecolor{mygray}{gray}{0.4}
\definecolor{autopurple}{HTML}{7030A0}
\definecolor{dyna_yellow}{HTML}{BF9000}
\definecolor{adaptive_blue}{HTML}{0070C0}
\definecolor{darksalmon}{rgb}{0.91, 0.59, 0.48}
\definecolor{emerald}{rgb}{0.31, 0.78, 0.47}
\definecolor{green(pigment)}{rgb}{0.0, 0.65, 0.31}
\definecolor{amaranth}{rgb}{0.9, 0.17, 0.31}
\definecolor{iris}{rgb}{0.35, 0.31, 0.81}
\definecolor{uu}{rgb}{0.95, 0.51, 0.51}
\definecolor{spirodiscoball}{rgb}{0.06, 0.75, 0.99}
\usepackage{svg}

\newcommand{\blue}[1]{$_{\color{BlueGreen}\downarrow #1}$}
\newcommand{\red}[1]{$_{\color{RedOrange}\uparrow #1}$}

\usepackage{cleveref}
\SetKwInOut{Input}{Input}\SetKwInOut{Output}{Output}

\SetCommentSty{mycommfont}
\SetKwComment{Comment}{$\triangleright$\ }{}

\usepackage{xspace}
\newcommand{\ourmethod}{{\fontfamily{lmtt}\selectfont \textbf{AgentPrune}}\xspace}
\newcommand{\llmname}[1]{{\fontfamily{pcr}\selectfont {#1}}\xspace}

\definecolor{ada_blue}{rgb}{0,205,205}
\definecolor{glt_red}{rgb}{109,205,255}
\definecolor{MorandiBlue}{RGB}{118,134,146}

\definecolor{demphcolor}{RGB}{144,144,144}
\definecolor{mygray}{gray}{0.4}
\definecolor{autopurple}{HTML}{7030A0}
\definecolor{dyna_yellow}{HTML}{BF9000}
\definecolor{adaptive_blue}{HTML}{0070C0}
\definecolor{darkgrey}{RGB}{120,120,120}
\definecolor{mygrey}{RGB}{200,200,200}

\usepackage[font=small,labelfont=bf]{caption}  
\usepackage{makecell}
\usepackage{tabulary}

\definecolor{myblue}{HTML}{00CDCD}
\definecolor{champagne}{rgb}{0.74, 0.83, 0.9}
\definecolor{champagne}{rgb}{0.97, 0.91, 0.81}

\newtheorem{definition}{Definition}

\hypersetup{
    colorlinks=true,
    linkcolor=red,
    citecolor=cyan,
    filecolor=magenta,      
    urlcolor=magenta,
    }

\title{Cut the Crap: An Economical Communication Pipeline for LLM-based Multi-Agent Systems}

\author{Guibin Zhang$^{1,2\dag}$, 
Yanwei Yue$^{1\dag}$, 
Zhixun Li$^{3\dag}$, 
Sukwon Yun$^{4}$, 
Guancheng Wan$^{5}$, \\
\textbf{Kun Wang}$^{6}$\thanks{Kun Wang is the corresponding author, $\dag$ denotes equal contributions.},\quad
\textbf{Dawei Cheng}$^{1,2}$,\;
\textbf{Jeffrey Xu Yu}$^{3}$,\;
\textbf{Tianlong Chen}$^{4}$
	\\
	$^{1}$Tongji University\quad
    $^{2}$Shanghai AI Laboratory \quad
    $^{3}$The Chinese University of Hong Kong \\
    $^{4}$University of North Carolina at Chapel Hill\quad
    $^{5}$Emory University \\
    $^{6}$Nanyang Technological University 
}

\iclrfinalcopy 
\begin{document}

\maketitle
\vspace{-0.5em}
\begin{abstract}
\vspace{-0.6em}
Recent advancements in large language model (LLM)-powered agents have shown that collective intelligence can significantly outperform individual capabilities, largely attributed to the meticulously designed inter-agent communication topologies. Though impressive in performance, existing multi-agent pipelines inherently introduce substantial token overhead, as well as increased economic costs, which pose challenges for their large-scale deployments. In response to this challenge, we propose an economical, simple, and robust multi-agent communication framework, termed \ourmethod, which can seamlessly integrate into mainstream multi-agent systems and prunes redundant or even malicious communication messages. Technically, \ourmethod is the first to identify and formally define the \textit{communication redundancy} issue present in current LLM-based multi-agent pipelines, and efficiently performs one-shot pruning on the spatial-temporal message-passing graph, yielding a token-economic and high-performing communication topology.
Extensive experiments across six benchmarks demonstrate that \ourmethod \textbf{(I)} achieves comparable results as state-of-the-art topologies at merely $\$5.6$ cost compared to their $\$43.7$, \textbf{(II)} integrates seamlessly into existing multi-agent frameworks with $28.1\%\sim72.8\%\downarrow$ token reduction, and \textbf{(III)} successfully defend against two types of agent-based adversarial attacks with $3.5\%\sim10.8\%\uparrow$ performance boost. The source code is available at \url{https://github.com/yanweiyue/AgentPrune}.
\end{abstract}

\vspace{-0.5em}
\section{Introduction}\label{sec:intro}
\begin{wrapfigure}{r}{0.55\textwidth}\vspace{-1.5em}
 \centering
 \includegraphics[width=\linewidth]{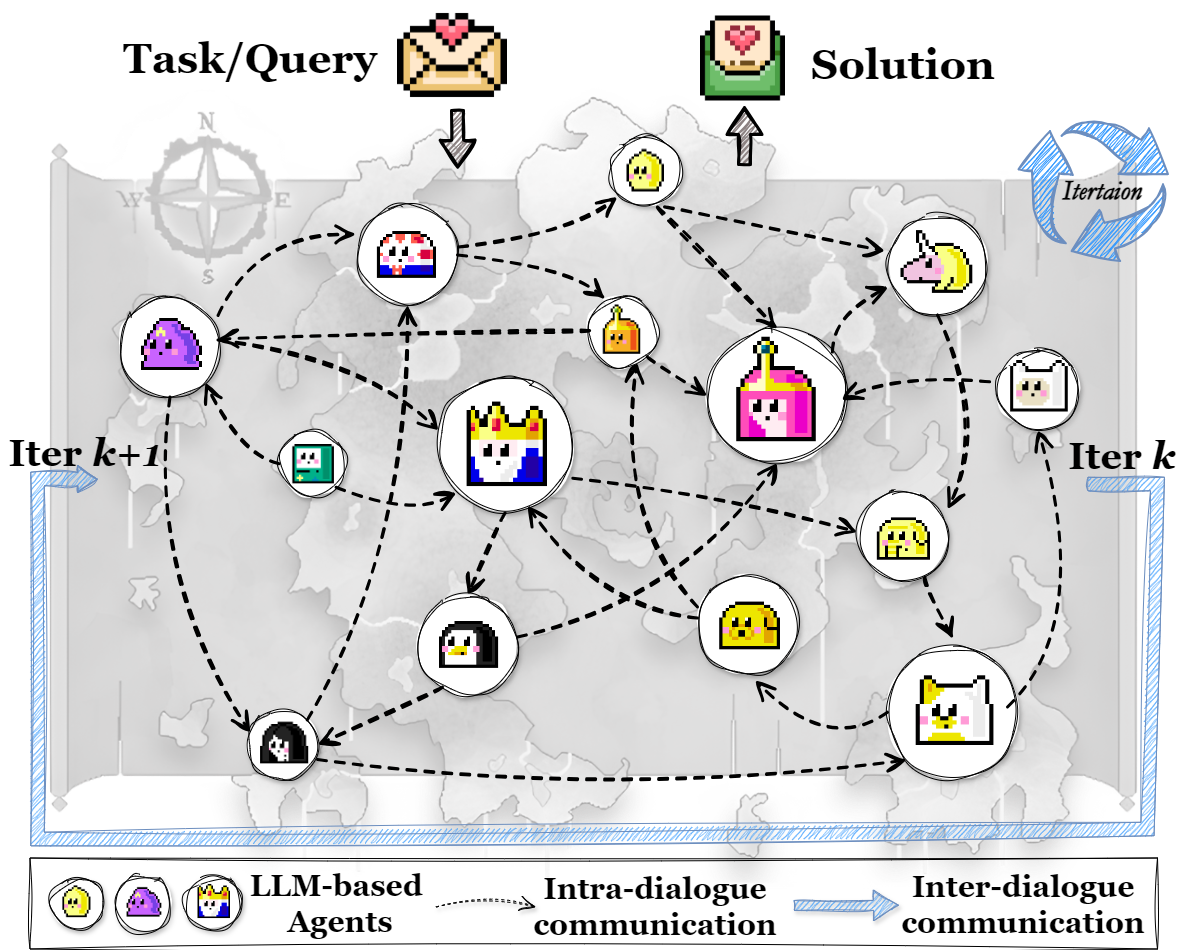}
  \vspace{-1.6em}
  \caption{The workflow of existing LLM-based (problem-solving) multi-agent systems. The agent-agent communication occurs at both intra- and inter-dialogue stages. \vspace{-1.4em} }
  \label{fig:intro}
\end{wrapfigure}
\vspace{-0.7em}
Large Language Model (LLM) based agents \citep{autogpt,babyagi,agentgpt} have demonstrated strong performance across a diverse range of tasks, including reasoning \citep{yao2023react}, code generation \citep{reflexion}, and even more complex applications like video gaming \citep{voyager} and autopilot systems \citep{jin2023surrealdriver}. Recent endeavors have shown that combining implicitly or explicitly different LLM-based agents into a team can outperform a single agent in handling complex tasks \citep{arXiv2023_MultiAgent-Debate, arXiv2023_MultiAgent-Debate_2, multi-persona, blender, reflexion, PHPrompting, autogen}, which supports the presence of human-esque collaborative intelligence in multi-agent systems~\citep{zhang2023exploring}. In practice, previous research
has explored approaches in which instances of LLMs, referred to as agents~\citep{FCS2024_Survey-Agent,arXiv2023_Survey-Agent_2,arXiv2023_Survey-Agent_3,arXiv2024_Survey-Agent_4,arXiv2024_Survey-Agents-CompExp}, collaborate synergistically (\textit{e.g.}, through debate or reflection) to complete tasks~\citep{arXiv2023_Survey-MultiAgentCooperation,arXiv2024_Survey-MultiAgent_2,arXiv2024_Survey-MultiAgent,arXiv2024_Survey-MultiAgent-System,arXiv2024_Survey-MultiAgent-System_2} via diverse communication topologies (\textit{e.g.}, chain~\citep{cot}, tree~\citep{tot}, complete graph~\citep{qian2024scaling}, random graph~\citep{qian2024scaling}, optimizable graph~\citep{zhuge2024gptswarm}, LLM-based network~\citep{chatllm-network,arXiv2023_Dynamic-LLM-Agent}). 
The exceptional performance of these cooperative agents significantly benefits from their interactive communication and collaboration, specifically how agents \textit{transmit, exchange,} and \textit{assimilate} information~\citep{chateval,multi-persona,arXiv2023_Dynamic-LLM-Agent}.

\begin{figure*}[]
\centering
\includegraphics[width=\linewidth]{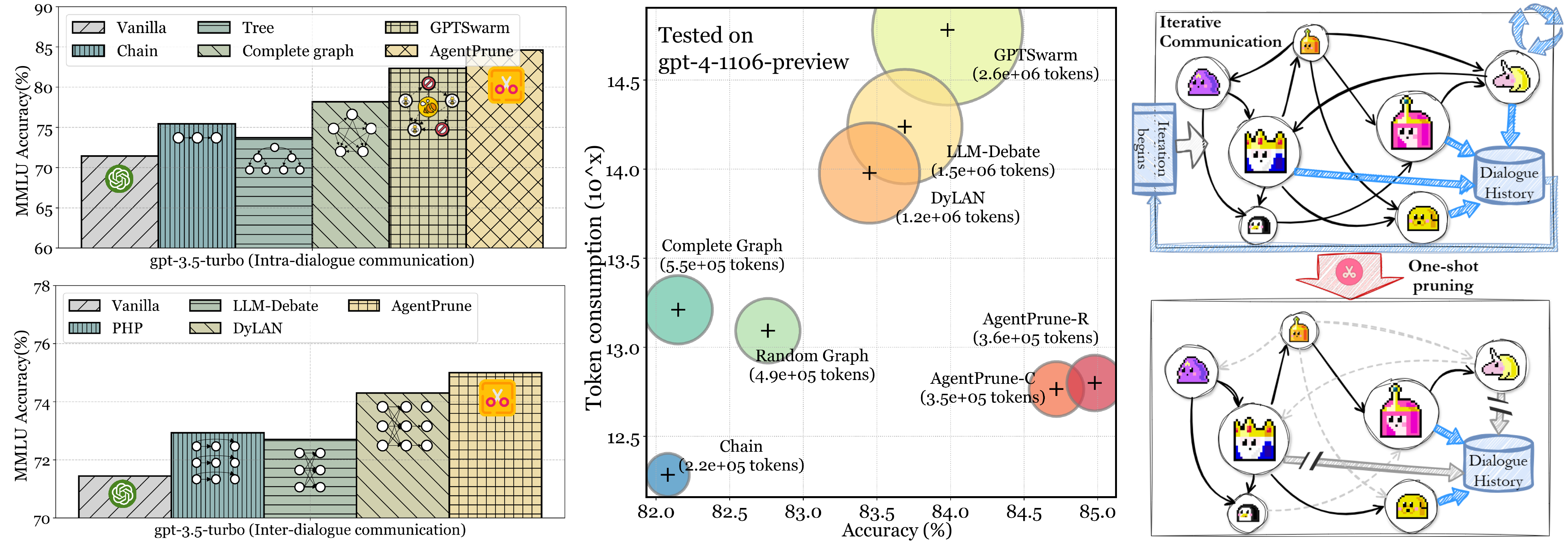}
\vspace{-6mm}
\caption{(\textbf{\textit{Left}}) The accuracy comparison on MMLU~\citep{mmlu} among (1) a single \llmname{gpt-3.5-turbo}, (2) three \llmname{gpt-3.5-turbo} as agents equipped with intra-dialogue communication structures like chain and tree~\citep{qian2024scaling}, complete graph, and GPTSwarm~\citep{zhuge2024gptswarm}, and (3) those equipped with inter-dialogue communication structures like PHP~\citep{PHPrompting}, LLM-Debate~\citep{arXiv2023_MultiAgent-Debate}, DyLAN~\citep{arXiv2023_Dynamic-LLM-Agent} and our \ourmethod. (\textbf{\textit{Middle}}) The prompt token consumption comparison on MMLU between different methods. (\textbf{\textit{Right}}) The overview of our proposed \ourmethod.}
\label{fig:intro_2}
\vspace{-2.2em}
\end{figure*}

Taking a closer look into the communication mechanisms in existing multi-agent systems, they typically involve two key types (as shown in \Cref{fig:intro}):
    \ding{182} \textbf{Intra-dialogue communication}: For a given query/task, multiple agents interact--whether by cooperating~\citep{arXiv2023_Survey-MultiAgentCooperation,autogen,rasal2024llm}, teaching~\citep{zhang2024classroom}, or competing~\citep{zhao2023competeai,bargaining-feedback}--to produce a solution within a dialogue round;
    \ding{183} \textbf{Inter-dialogue communication}: In a specific manner—whether by summarizing~\citep{chateval,shen2024smallllms}, replicating~\citep{yin2023exchange,arXiv2023_MultiAgent-Debate}, or filtering~\citep{arXiv2023_Dynamic-LLM-Agent}—the content of the current dialogue is passed to the next round of interaction as a reference, initiating a new cycle of collaborative efforts. 

To better illustrate the power of agent communication, \Cref{fig:intro_2} (\textit{Left}) compares the performance of a single \llmname{gpt-3.5-turbo} with three agents equipped with different inter/intra-dialogue communication structures. The results demonstrate that even the simplest communication framework significantly leads to a notable accuracy improvement, which vividly showcases the social intelligence and collaborative capabilities of LLMs~\citep{PNAS2024_TuringTest_Chatbots-Humans}. 
However, the success of multi-agents comes at the cost of significantly increased token consumption, imposing substantial economic burdens~\citep{wang2024tokeneconomy}, which are detrimental to the widespread application of multi-agent systems, as deployment on edge devices does not accommodate excessively costly inference~\citep{liu2023deja}. A piece of empirical evidence is in \Cref{fig:intro_2} (\textit{Middle}), where various communication methods result in a $2\sim11.8\times$ increase in token consumption compared to the simple chain structure, severely undermining the \textbf{token economy} of existing multi-agent systems. 

In the light of this limitation, we \textit{for the first time} identify a significant phenomenon of \textit{Communication Redundancy}~\citep{meyer2021entspann} within existing LLM-based multi-agent (LLM-MA) communication topologies, where a substantial portion of message passing does not contribute meaningfully to the collaborative intelligence. With this finding, we introduce \textit{an economical and versatile communication pruning framework for LLM-powered multi-agent systems}, dubbed \ourmethod, which can be smoothly incorporated within various existing LLM-MA systems, offering comparable reasoning and planning performance as well as significantly lower token consumption. Practically, \ourmethod treats the entire LLM-MA framework as a spatial-temporal communication graph, in which each agent, along with its unique properties (\textit{e.g.}, profile~\citep{NeurIPS2023_Agent-SoM}, external API tools~\citep{zhuang2023toolchain}, or knowledge base~\citep{chen2024benchmarking}), is packaged as a node, communication between agents within the same dialogue forms spatial edges, and communication across dialogues forms temporal edges. By training a low-rank-principle-guided graph mask, \ourmethod efficiently identifies the important graph connectivities (\textit{i.e.}, message passing through edges). This comes with a one-shot pruning to derive a sparse yet informative communication graph (in \Cref{fig:intro_2} (\textit{Right})), which is then fixed as the communication topology for subsequent token-economic and efficient reasoning.
Our contributions can be summarized as follows:

\vspace{-0.7em}
\begin{itemize}[leftmargin=*]
    \item[\ding{182}] \textbf{\textit{System Discovery.}} We present a spatial-temporal graph paradigm to describe the communication topology of contemporary LLM-MA frameworks, and further identify and define the \textit{Communication Redundancy} issue in current systems, wherein a significant portion of spatial and temporal edges, \textit{i.e.}, communication, does not contribute to collaborative intelligence.

    \item[\ding{183}] \textbf{\textit{Pratical Solution.}} We propose \ourmethod, an economical, simple, and robust multi-agent communication pruning pipeline. By leveraging a trainable communication graph mask, \ourmethod identifies key message exchanges and prunes non-essential components in a one-shot manner, resulting in a sparse, token-economical, and highly informative communication graph. Notably, \ourmethod employs a low-rank principle to guide the graph mask training, successfully robustifying LLM-MA systems against various agent-targeted adversarial attacks.

    \item[\ding{184}] \textbf{\textit{Experimental Validation.}} Extensive experiments on six benchmarks show that \ourmethod is: \textbf{(1) high-performing}, achieving comparable performance on MMLU at $\$5.6$ cost, to that of state-of-the-art communication topologies at $\$43.7$; \textbf{(2) token-economical}, integrating seamlessly into popular multi-agent frameworks including AutoGen and GPTSwarm, reducing their token cost by $28.1\%\sim72.8\%\downarrow$; and \textbf{(3) adversarially robust}, successfully defending against two types of agent adversarial attacks, with a $3.5\%\sim10.8\%\uparrow$ performance improvement.
\end{itemize}
\vspace{-0.5em}

\vspace{-0.5em}
\section{LLM-MA as Spatial-temporal Graphs}
\vspace{-0.5em}
\paragraph{Notations} We describe the whole multi-agent system as a graph $\mathcal{G}=(\mathcal{V},\mathcal{E})$, with $\mathcal{V}=\{v_1,v_2,\cdots,v_{|\mathcal{V}|}\}$ being the node set and $\mathcal{E}$ being the edge set. Each node $v_i\in \mathcal{V}$ represents an agent, which can be further interpreted as follows:
\vspace{-0.3em}
\begin{equation}
v_i = \{\texttt{Base}_i, \texttt{Role}_i, \texttt{State}_i, \texttt{Plugins}_i\},\;\;\texttt{Plugins}_i=\{\texttt{F}_j,\texttt{C}_j\}_{j=1}^{P} ,
\vspace{-0.2em}
\end{equation}
where an agent $v_i$ consists of: (1) $\texttt{Base}_i$, the language model instance used by $v_i$;
(2) $\texttt{Role}_i$, the pre-defined role or responsibility of the agent;
(3) $\texttt{State}_i$, the state of the agent, encapsulating the accumulated knowledge and experience from previous interactions; (4) $\texttt{Plugins}_i$, a set of $P$ external plugins available to agent $v_i$, where each plugin is defined by its functionalities $\texttt{F}_j$ (\textit{e.g.}, web search, python compiler) and configurations $\texttt{C}_j$. 
\vspace{-0.4em}

For the edges, we divide them into two subsets: \textbf{intra-dialogue (spatial) edges} $\mathcal{E}^{\mathcal{S}}\subseteq\mathcal{V}^{(t)}\times\mathcal{V}^{(t)}$ and \textbf{inter-dialogue (temporal) edges} $\mathcal{E}^{\mathcal{T}}\subseteq \mathcal{V}^{(t-1)}\times\mathcal{V}^{(t)}$. For each spatial edge $e^{\mathcal{S}}_{ij} = (\mathbf{M}_{ij}, \mathbf{O}_{ij})$, it represents the information flow from agent $v_i$ to agent $v_j$ \textit{within the same utterance}, composed of the message content $\mathbf{M}_{ij}$, as well as the associated operation $\mathbf{O}_{ij}$ (\textit{e.g.}, task assignments, requests). For each temporal edge $e^{\mathcal{T}}_{ij}$, it denotes the message passing \textit{between two utterance rounds}, \textit{i.e.}, whether the output from agent $v_i$ in the $(t-1)$-th round should be passed on to agent $v_j$ in the $t$-th round. Further, we define the temporal/spatial (in-)neighbors for each agent as follows:
\vspace{-0.3em}
\begin{equation}
\mathcal{N}^\mathcal{T}(v_i) = \{v_j\;|\;(j,i)\in\mathcal{E}^\mathcal{T}\},\;\mathcal{N}^\mathcal{S}(v_i) = \{v_j\;|\;(j,i)\in\mathcal{E}^\mathcal{S}\}.
\vspace{-0.5em}
\end{equation}
\paragraph{Multi-agent Communication}\label{sec:multi-agent-graph} We provide a graph-based description of the reasoning process in task-oriented multi-agent systems. Given a query/task $q$, it is sequentially fed to each agent, which produces its output. 
To maintain an orderly sequence of agent interactions, we utilize topological ordering~\citep{bondy1976graph} to ensure that each node is processed only after all its dependencies have been addressed. This necessitates that the spatial communication graph $\mathcal{G}^{\mathcal{S}} = (\mathcal{V}, \mathcal{E}^{\mathcal{S}})$ be structured as a directed acyclic graph (DAG).
Formally, for $\mathcal{G}^{\mathcal{S}}$, the following condition holds:
\begin{equation}
\forall (v_i ,v_j),\;\mathbb{I}(v_i) < \mathbb{I}(e_{ij}) < \mathbb{I} (v_j),
\end{equation}
where $\mathbb{I}(x)$ denotes the execution order of $x$. For each agent $v^{(t)}_i$ at round $t$, it produces its rationale or answers, uniformly denoted as $\mathbf{M}_{i}$, as follows:\vspace{-0.3em}
\begin{equation}\label{eq:agent_action}
\mathbf{M}^{(t)}_i \sim \mathcal{P}_\theta\Biggl(\mathbf{M}_i\;|\;q,\; \texttt{Role}^{(t)}_i,\; \texttt{State}^{(t)}_i, \overbrace{\cup_{\scriptscriptstyle v_j \in \mathcal{N}^{\mathcal{T}}(v_i)}\!\!\!\!\!\mathbf{M}_{j}}^{{\text{temporal}}}, \overbrace{\cup_{\scriptscriptstyle v_j \in \mathcal{N}^{\mathcal{S}}(v_i)}\!\!\!\!\mathbf{M}_{ji}}^{{\text{spatial}} }\Biggl),
\vspace{-0.8em}
\end{equation}
where agent $v_i$ responds based on the query $q$, its current role and state, temporal and spatial messages, and certain prompting instruction $\mathcal{P}_\theta$. Typically, after $K$ rounds of dialogue, a summarizer agent or an answer aggregation mechanism (\textit{e.g.}, voting) is employed to produce the final solution $a^{(K)}$ for the given query $q$. We conclude the general pipeline in \Cref{algo:llmma}.

\clearpage
\vspace{-1.4em}
\begin{algorithm}[!ht]\small
\DontPrintSemicolon
\SetAlgoLined
\LinesNumbered
\SetKwFunction{Fstop}{MeetEndCondition()}
\SetKwFunction{Ftopo}{TopologicalSort}
\SetKwFunction{Faggregate}{AggregateSolution}
\KwIn{Query $q$, Communication graph $\mathcal{G}=\{\mathcal{G}^\mathcal{S},\mathcal{G}^\mathcal{T}\}$, Maximum number of iterations $N$}

\For{\rm{iteration} $t\leftarrow1$ \KwTo $N$}{
\label{algo:line_start}
\If{\Fstop}{break \tcp{Extra stopping criteria like agent consensus}}
\For{$v_i$ \rm{in} $\Ftopo(\mathcal{V})$}{

$\boldsymbol{m}^\mathcal{T}\gets \{\mathbf{M}_j\;|\;v_j\in\mathcal{N}^\mathcal{T}(v_i)\}$ 
 \tcp{Messages from temporal in-neighbors}

$\boldsymbol{m}^\mathcal{S} \gets \{\mathbf{M}_{ij}\;|\;v_j\in\mathcal{N}^\mathcal{S}(v_i)\}$ 
 \tcp{Messages from spatial in-neighbors}

$\mathbf{M}_i^{(t)} \sim \mathcal{P}_{\theta} (\mathbf{M}_i\;|\;q, \texttt{Role}^{(t)}_i,\texttt{State}^{(t)}_i, \boldsymbol{m}^\mathcal{T}, \boldsymbol{m}^\mathcal{S})$ 
  \tcp{Generate rationale or answer}

}
$a^{(t)} \gets \Faggregate(\mathbf{M}_1^{(t)}, \mathbf{M}_2^{(t)},\cdots, \mathbf{M}_{|\mathcal{V}|}^{(t)})$ \tcp{Depending on the specific system, possible implementations of \texttt{AggregateSolution} include (but are not limited to) majority voting or using the output of a summarizer agent.}
\label{algo:line_end}
}
\Return $a^{(t)}$ as the final solution
\caption{Execution pipeline of LLM-MA systems from spatial-temporal graph perspective}
\label{algo:llmma}
\end{algorithm}

\vspace{-2em}
\paragraph{Problem Formulation}
\begin{wrapfigure}{r}{0.55\textwidth} \vspace{-1.5em}
 \centering
 \includegraphics[width=\linewidth]{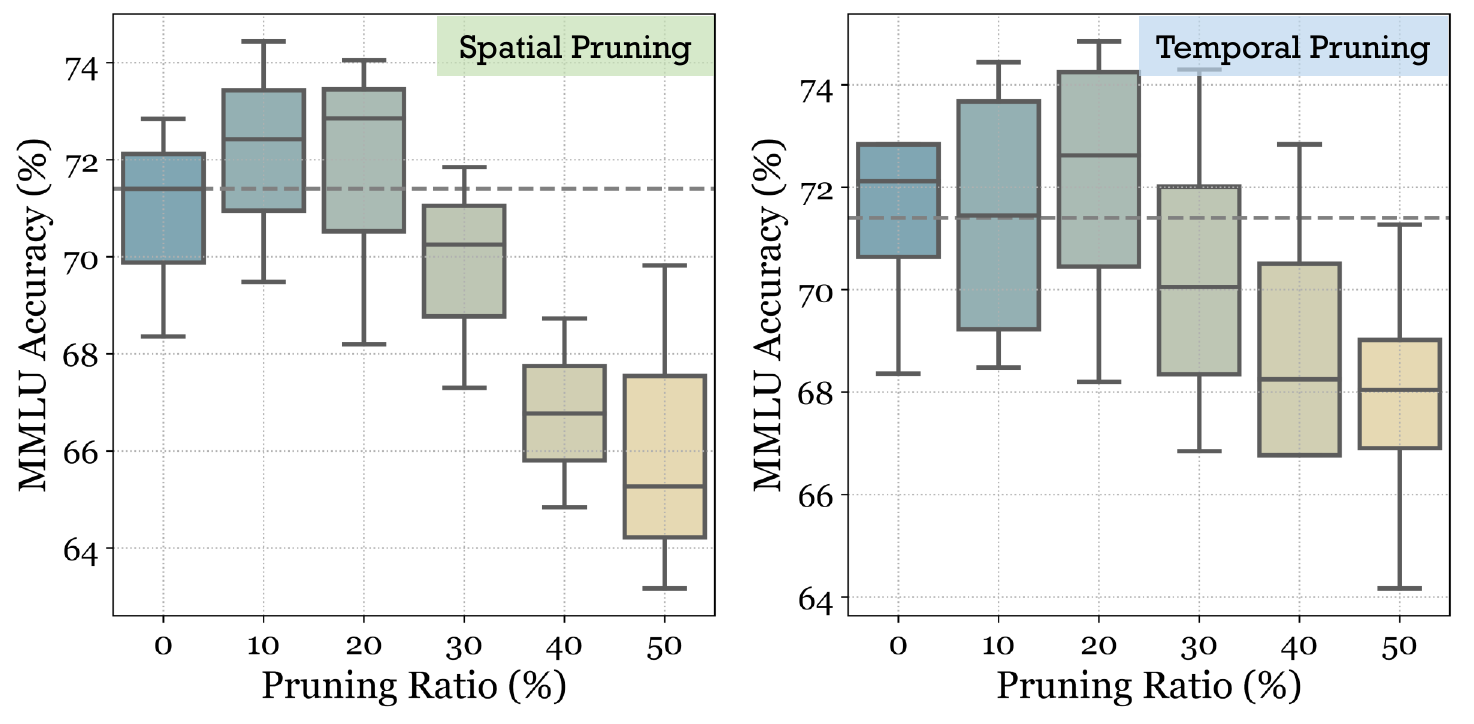}
  \vspace{-1.9em}
  \caption{The performance of mesh graph and LLM-Debate structure under different random pruning ratios on MMLU. \vspace{-1.5em} }
  \label{fig:preliminary}
\end{wrapfigure}
In this section, 
we explore and define the \textit{communication redundancy} issue within existing multi-agent communication pipelines. Specifically, we examine two representative communication topologies: (1) for \textit{spatial communication}, the fully-connected mesh graph from MacNet~\citep{qian2024scaling}, which exemplifies a densely structured intra-utterance communication, and (2) for \textit{temporal communication}, the LLM-Debate~\citep{arXiv2023_MultiAgent-Debate}, where at the start of each dialogue round, an agent receives all responses from the previous round as input. Using four \llmname{gpt-3.5-turbo} as agents, we assess system performance on MMLU after randomly pruning a certain proportion of connections. As illustrated in \Cref{fig:preliminary}, when randomly removing $10\%\sim30\%$ of the communication connectivity, the performance actually gains up to $2.83\%$ improvement. This suggests that, in both spatial and temporal information flow, a substantial portion of messages does not contribute to the task-solving process, which we formally define as follows:
\begin{definition}[Communication Redundancy]
For any LLM-based multi-agent communication graph $\mathcal{G} = (\mathcal{V}, \mathcal{E}^\mathcal{S} \cup \mathcal{E}^\mathcal{T})$, the following condition holds:
\begin{equation}
\exists\; \mathcal{G}^\text{sub} = (\mathcal{V}, \mathcal{E}' \cup \mathcal{E}'') \subseteq \mathcal{G}, \; \operatorname{where} \; \mathcal{E}' \subseteq \mathcal{E}^\mathcal{S}, \; \mathcal{E}'' \subseteq \mathcal{E}^\mathcal{T}, \; \operatorname{s.t.} \; \phi(\mathcal{G}^\text{sub}) \geq \phi(\mathcal{G}),
\end{equation}
where $\phi(\cdot)$ represents a utility function that measures the solution quality achieved by the system. The redundant components in the communication topology, denoted as $(\mathcal{E}^{st} \setminus \mathcal{E}') \cup (\mathcal{E}^{tp} \setminus \mathcal{E}'')$, are referred to as the communication redundancy in LLM-MA systems.
\end{definition}
\vspace{-0.6em}
We further outline the objective of this study as follows:
\begin{equation}\label{eq:problem_definition}
{\arg \max}_{\mathcal{E}',\mathcal{E}''} \mathcal{G} \setminus \mathcal{G}^\text{sub},\;\text{s.t.} \; |\phi(\mathcal{G}^\text{sub}) - \phi(\mathcal{G})| \leq \epsilon,
\end{equation}
where $\epsilon$ represents the allowable threshold for performance variation. \Cref{eq:problem_definition} aims to minimize communication redundancy with performance guarantee.

\vspace{-0.9em}
\section{Methodology}
\begin{figure*}[!t]
\centering
\includegraphics[width=\linewidth]{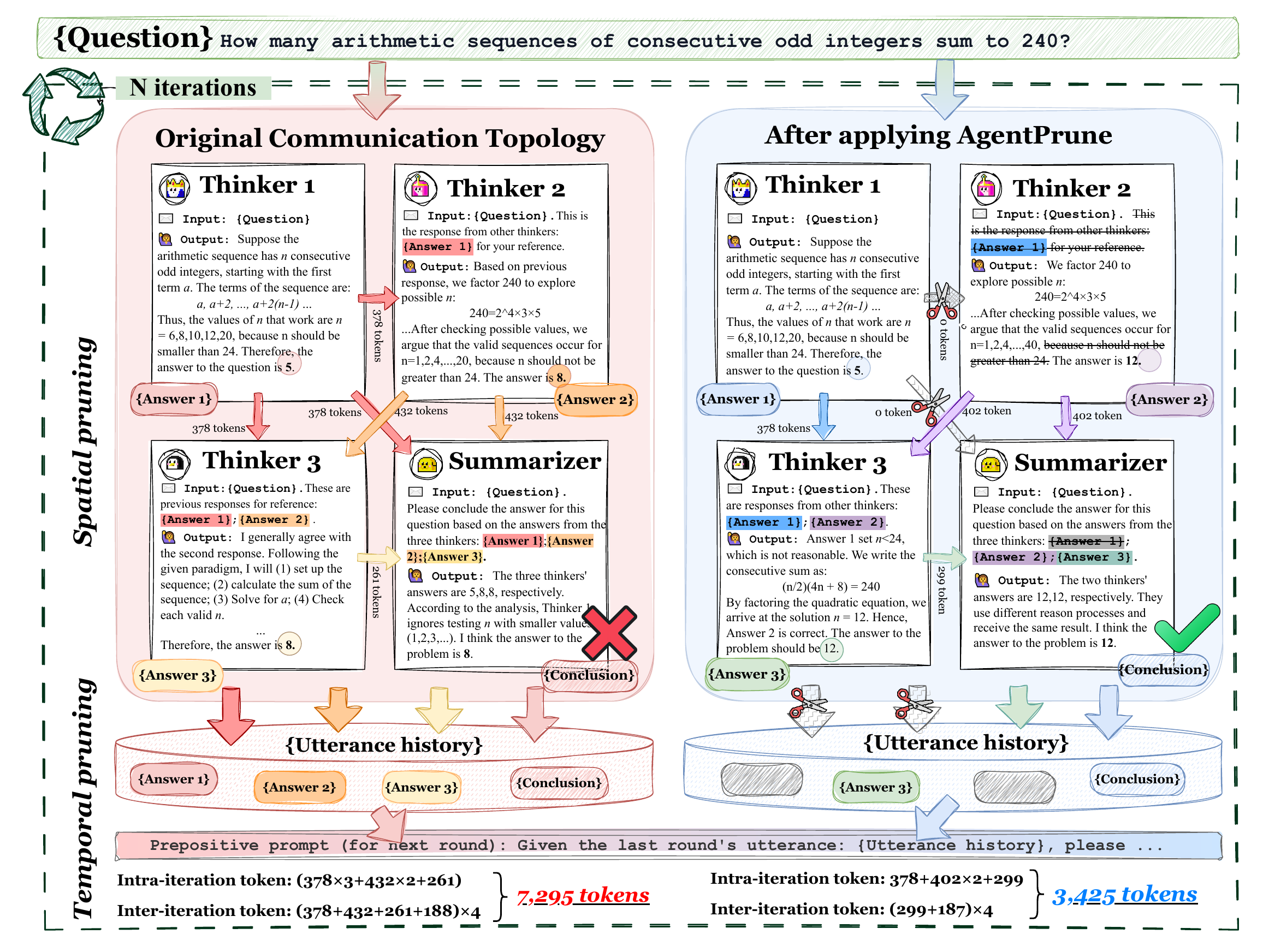}
\vspace{-2em}
\caption{The overview of our proposed \ourmethod.}
\label{fig:framework}
\vspace{-1.3em}
\end{figure*}

\vspace{-0.8em}
\Cref{fig:framework} illustrates how our method is applied within an LLM-MA system. Specifically, given an input query, \ourmethod first performs \textit{spatial pruning} by eliminating redundant spatial messages within a dialogue round, followed by \textit{temporal pruning} to discard unnecessary dialogue history. 
In the following sections, we will first explain how \ourmethod facilitates efficient multi-round communication based on an optimizable spatial-temporal communication graph ($\vartriangleright$ \Cref{sec:stgraph,sec:optimize_conn}), leverages one-shot pruning to derive a sparse interaction topology ($\vartriangleright$ \Cref{sec:one-shot}), and finally,  detail the optimization paradigm for the entire framework ($\vartriangleright$ \Cref{sec:opt}).

\vspace{-1em}
\subsection{Spatial-temporal Graph Communication}\label{sec:stgraph}
\vspace{-0.5em}

Given an arbitrary LLM-MA system and its corresponding spatial-temporal communication graph $\mathcal{G}$, the task of \ourmethod is to discover its sparse yet equally high-performing counterpart $\mathcal{G}^{\text{sub}}$. The objective is essentially a graph sparsification problem~\citep{spielman2008graph,chen2023demystifying}, whose goal is to identify the essential graph connections and discard the less critical ones. 
To achieve this, following classical practices in graph sparsification~\citep{chen2021unified,wang2023brave,zhang2024graph}, we relax the original binary communication graph $\mathcal{G}$ by transforming its edge elements from binary values to continuous variables, denoted as $\Tilde{\mathcal{G}}$. We have: 
\begin{equation}
\mathbf{A}({\mathcal{G}}) = \{\mathbf{A}^\mathcal{S}, \mathbf{A}^\mathcal{T}\},\;\mathbf{A}(\Tilde{\mathcal{G}}) = \mathbf{A}(\{\Tilde{\mathcal{G}}^\mathcal{S},\Tilde{\mathcal{G}}^\mathcal{T}\}) = \{\mathbf{A}^\mathcal{S}\odot \mathbf{S}^\mathcal{S}, \mathbf{A}^\mathcal{T}\odot \mathbf{S}^\mathcal{T}\},
\end{equation}
where $\mathbf{A}(\mathcal{G})$ obtains the adjacency matrix of input graph $\mathcal{G}$, and $\mathbf{A}^{\mathcal{S}}, \mathbf{A}^{\mathcal{T}}\in \{0,1\}^{|\mathcal{V}|\times|\mathcal{V}|}$ represent the spatial and temporal adjacency matrices, respectively. Specifically, $\mathbf{A}^{x}[i,j] = 1$ indicates that $e_{ij} \in \mathcal{E}^{x}$, and $0$ otherwise. It is important to note that both $\mathbf{A}^{\mathcal{S}}$ and $\mathbf{A}^{\mathcal{T}}$ are predefined by the LLM-MA system. $\mathbf{S}^\mathcal{S}, \mathbf{S}^\mathcal{T}\in\mathbb{R}^{|\mathcal{V}|\times|\mathcal{V}|}$ are differentiable graph masks. As mentioned in \Cref{sec:multi-agent-graph}, we require the interaction topology to be a DAG to ensure that agent input/output (I/O) can be processed sequentially. Therefore, we leverage a \texttt{DAGSampling} function to transform the original $\Tilde{\mathcal{G}}^\mathcal{S}$ into a DAG: $\hat{\mathcal{G}}^\mathcal{S}\leftarrow \texttt{DAGSampling}(\Tilde{\mathcal{G}}^\mathcal{S})$, whose procedure is described in \Cref{algo:dag}. 
\vspace{-0.8em}
\subsection{Optimizing Spatial-temporal Connectivity}\label{sec:optimize_conn}
\vspace{-0.8em}
With $\hat{\mathcal{G}}^{\mathcal{S}}$ and $\Tilde{\mathcal{G}}^\mathcal{T}$ in hand, we aim to optimize them toward both high-performance and token efficiency. To this end, we introduce two optimization objectives for $\hat{\mathcal{G}}^{\mathcal{S}}$ and $\Tilde{\mathcal{G}}^\mathcal{T}$: \ding{182} \textit{distribution approximation}, ensuring accurate estimation of their underlying probability distributions, and \ding{183} \textit{low-rank sparsity}, which promotes a more efficient and sparse structure \citep{li2024gslb}. The first objective ensures that the magnitudes of the graph masks correctly reflect the importance of different communication channels, facilitating subsequent redundancy pruning, and the second ensures that the learned connectivity remains sparse and robust. Formally, we define the following objective:
\vspace{-0.6em}
\begin{equation}
\label{eq:objective}
\underset{\mathbf{S}^\mathcal{S},\mathbf{S}^\mathcal{T}\in \mathbb{S}}{\arg \max}\; \overbrace{\mathbb{E}_{\hat{\mathcal{G}}^{\mathcal{S}},\mathcal{G}^\mathcal{T}\sim\mathbb{G}}\left[\phi\left(\{\hat{\mathcal{G}}^{\mathcal{S}},\Tilde{\mathcal{G}}^\mathcal{T}\}\right)\right]}^\text{distribution approximation}-\!\!\overbrace{\sum_{\scriptscriptstyle \mathcal{X}\in\{\mathcal{S},\mathcal{T}\}}\!\!\operatorname{rank}(\mathbf{S}^\mathcal{X})}^\text{low-rank sparsity},\; \operatorname{s.t.}\!\!\!\!\sum_{\scriptscriptstyle \mathcal{X}\in\{\mathcal{S},\mathcal{T}\}}\!\!\!||\mathbf{A}^\mathcal{X} - \mathbf{S}^\mathcal{X}||_F\leq \delta,
\end{equation}
where $\mathbb{S}$ and $\mathbb{G}$ represent the viable parameter space, $\phi(\cdot)$ serves as the utility evaluator for the input multi-agent framework, $\operatorname{rank}(\cdot)$ calculates the rank of matrix, and $\delta$ is the noise level. Next, we will provide a detailed explanation of the implementation of these two optimization objectives.
\vspace{-0.6em}
\paragraph{Distribution Approximation} The first term in \Cref{eq:objective} encourages $\{\mathbf{S}^\mathcal{S},\mathbf{S}^\mathcal{T}\}$ towards the maximization of the system's utility. However, since $\phi(\cdot)$ often depends on external APIs~\citep{li2023api} or compilers~\citep{human-eval} for evaluation, it is generally non-differentiable. Therefore, we employ policy gradient~\citep{williams1992simple} to make \Cref{eq:objective} tractable:
\begin{gather}
\label{eq:policy}
\nabla_{\mathbf{S}}\; \mathbb{E}_{\hat{\mathcal{G}}^{\mathcal{S}},\Tilde{\mathcal{G}}^\mathcal{T}\sim\mathbb{G}}\left[\phi\left(\{\hat{\mathcal{G}}^{\mathcal{S}},\Tilde{\mathcal{G}}^\mathcal{T}\}\right)\right] \approx \frac{1}{M}\sum_{k=1}^M\phi\left(\{\hat{\mathcal{G}}^\mathcal{S}_k, \Tilde{\mathcal{G}}^\mathcal{T}_k\}\right)\nabla_{\mathbf{S}}\log \left(p_{\mathbf{S}}(\{\hat{\mathcal{G}}^\mathcal{S}_k, \Tilde{\mathcal{G}}^\mathcal{T}_k\})\right),\\
p_\mathbf{S}\left(\{\hat{\mathcal{G}}^\mathcal{S}_k, \Tilde{\mathcal{G}}^\mathcal{T}_k\}\right) = \left(\prod \mathbbm{1}_{e_{ij}\in \mathcal{E}^\mathcal{S}} \mathbf{S}^\mathcal{S}[i,j]\right) \cdot  \left(\prod \mathbbm{1}_{e_{ij}\in \mathcal{E}^\mathcal{T}} \mathbf{S}^\mathcal{T}[i,j]\right)
\end{gather}
where $\mathbf{S} = \{\mathbf{S}^\mathcal{S}, \mathbf{S}^\mathcal{T}\}$, $\{\hat{\mathcal{G}}^{\mathcal{S}}, \Tilde{\mathcal{G}}^{\mathcal{T}}\}_{k=1}^M$ are independently sampled from $\{\hat{\mathcal{G}}^{\mathcal{S}}, \Tilde{\mathcal{G}}^\mathcal{T}\}$,  $p_\mathbf{S}(\{\hat{\mathcal{G}}^\mathcal{S}_k, \Tilde{\mathcal{G}}^\mathcal{T}_k\})$ calculates the probability of the sampled structure, and $\mathbbm{1}(\cdot)$ is an indicator function. 
\vspace{-0.6em}
\paragraph{Low-rank Sparsity} The second term in \Cref{eq:objective} promotes the graph masks $\{\mathbf{S}^\mathcal{S},\mathbf{S}^\mathcal{T}\}$ to be low-rank, which not only filters out informative agent communications but also aids in removing redundant, noisy, and even malicious messages, which has been demonstrated in recent studies, showing that low-rank graphs are more robust to network attacks~\citep{entezari2020all,ennadir2024simple}. We will empirically validate \ourmethod's ability to enhance multi-agent robustness in \Cref{exp:robust}. However, directly optimizing the rank minimization is NP-hard, so we replace the rank function with the nuclear norm as an alternative, reformulating this term as follows:
\begin{equation}\label{eq:rank_1}
\underset{\mathbf{S}^\mathcal{S},\mathbf{S}^\mathcal{T}\in \mathbb{S}}{\arg \min} {\textstyle\sum}_{\scriptscriptstyle \mathcal{X}\in\{\mathcal{S},\mathcal{T}\}}||\mathbf{S}^\mathcal{X}||_*,\;\operatorname{s.t.} ||\mathbf{A}^\mathcal{X} - \mathbf{S}^\mathcal{X}||_F\leq \delta,
\end{equation}
where $||\mathbf{S}||_* = \sum_i \sigma_i$, and $\sigma_i$ represents the $i$-th singular value of $\mathbf{S}$. 
Guided by \Cref{eq:objective}, we iteratively optimize the spatial-temporal connectivity in conjunction with the multi-agent conversation over \(K'\) rounds, where \(K' \ll K\).

\vspace{-0.5em}
\subsection{One-shot Pruning}\label{sec:one-shot}
\vspace{-0.5em}
We dynamically optimize $\{\mathbf{S}^\mathcal{S}, \mathbf{S}^\mathcal{T}\}$ for only $K'$ iterations, rather than the full $K$ iterations, because prior work on Early-bird (EB) and Graph EB has demonstrated that limited training can also construct high-quality benchmarks reflecting the topology distribution~\citep{achille2018critical,you2019drawing,zhang2024two}, which also aligns with \ourmethod's token-saving initiation. To eliminate redundancy in the current communication structure, we perform one-shot magnitude pruning on the optimized graph masks $\mathbf{S}$ (either $\mathbf{S}^\mathcal{S}$ or  $\mathbf{S}^\mathcal{T}$):
\begin{equation}\label{eq:oneshot}
\mathbf{B} = \mathbbm{1}\biggl(\mathbf{A} \neq 0 \wedge \operatorname{TopK}\Bigl(\mathbf{S}, |\mathbf{A}|\times \left(1-p\%\right)\Bigl)\biggl),
\end{equation}
where $\operatorname{TopK}(S,x\%)$ return the largest $x\%$ elements in matrix $S$, and $p\%$ is the pruning ratio. By applying the binary masks to the original topology, we obtain sparse, compact, and communication-minimizing connectivity $\mathcal{G}^\text{sub}$, where $
\mathbf{A}(\mathcal{G}^\text{sub}) 
= \{\mathbf{A}^\mathcal{S} \odot \mathbf{B}^\mathcal{S}, \mathbf{A}^\mathcal{T}\odot \mathbf{B}^\mathcal{T}\}.$
In the subsequent $(K-K')$ rounds, the entire framework's message passing pipeline is strictly constrained by $\mathcal{G}^\text{sub}$, and agents are continuously optimized to refine the solution for query $q$.
\vspace{-0.5em}
\subsection{Application and Analysis}\label{sec:opt}
\vspace{-0.5em}
\paragraph{Algorithm Pipeline} As a plug-and-play module, \ourmethod can be harmoniously embedded in mainstream multi-agent frameworks to facilitate token-efficient communication, provided that the number of agents exceeds three and the communication structure is moderately organized (\textit{e.g.}, chain or direct-output structures are too simple to be applicable). When combined with \ourmethod, multiple agents first undergo $K'$ rounds of interactions alongside trainable graph masks, which are then one-shot pruned to yield the sparse $\mathcal{G}^\text{sub}$, leveraged for the subsequent $(K-K')$ rounds of optimization. We summarize all the notations used in \Cref{app:notation} and the comprehensive algorithmic workflow in \Cref{app:workflow}.
\vspace{-1em}
\paragraph{Multi-Query Training} For complex tasks like repository-level code generation~\citep{software-dev,liu2024evaluating}, multi-turn dialogues are often inevitable. However, for simpler tasks that involve a large number of queries, such as multiple choice answering~\citep{agashe2023evaluating,qian2024scaling}, typically only one or two dialogue rounds are needed, according to previous practices~\citep{yin2023exchange}. Under such circumstances, optimizing the connectivity for each query independently can be unnecessarily costly. Therefore, we give a \textit{multi-query training paradigm} for \ourmethod, which optimizes and prunes the spatial-temporal topology using merely $Q' (Q'\ll Q)$ queries, given a dataset composed of $Q$ queries.
See details in \Cref{app:multi_query}.
\vspace{-1em}
\paragraph{Cost Analysis}\label{sec:cost} In this section, we quantify the difference in token consumption between \ourmethod and the vanilla pipeline. Given a communication graph $\mathcal{G}$ and $K$ dialogue rounds, assuming that the average token count per spatial/temporal/query message is $c_\mathcal{S}, c_\mathcal{T}, c_q$, respectively, then the total token consumption of the vanilla system is $
C_\mathcal{G} = K\left[ c_\mathcal{S} |\mathcal{E}^\mathcal{S}| + c_\mathcal{T} |\mathcal{E}^\mathcal{T}| + C_q|\mathcal{V}| \right]$. The token consumption after applying \ourmethod is divided into two stages. The first stage involves $MK'\left[ c_\mathcal{S} |\mathcal{E}^\mathcal{S}| + c_\mathcal{T} |\mathcal{E}^\mathcal{T}| + C_q|\mathcal{V}| \right]$, while in the second stage, after the topology is fixed, the consumption becomes $(K-K')\left[ (1-p\%)\cdot\left(c_\mathcal{S} |\mathcal{E}^\mathcal{S}|+c_\mathcal{T} |\mathcal{E}^\mathcal{T}|\right) + C_q|\mathcal{V}| \right]$. Therefore, the total token savings $\Delta$ achieved by \ourmethod can be expressed as:
\vspace{-0.5em}
\begin{equation}
\Delta = \Bigl( \left(1+p\%\right)K - \left(M+p\%\right)K'  \Bigl)  \Bigl(c_\mathcal{S} |\mathcal{E}^\mathcal{S}|+c_\mathcal{T} |\mathcal{E}^\mathcal{T}| \Bigl) + (1-M)K' C_q|\mathcal{V}|. \vspace{-0.5em}
\end{equation}
We present the cost analysis of \ourmethod in multi-query training in \Cref{app:cost_analysis}. We will empirically evaluate the substantial token savings gained by \ourmethod in \Cref{exp:perf_cost}.

\vspace{-0.4em}
\section{Experiments}
\vspace{-0.8em}
In this section, we conduct extensive experiments to answer the following research questions: 
(\textbf{RQ1}) How does \ourmethod perform with respect to task completion and token efficiency?
(\textbf{RQ2}) Can \ourmethod reduce the economical cost of existing multi-agent systems without compromising performance?
(\textbf{RQ3}) Is \ourmethod effective in defending against adversarial attacks on agents?
(\textbf{RQ4}) How sensitive is \ourmethod to its key components or parameters?

\begin{table*}[!t]
\centering
\label{tab:main_results}
\caption{Performance comparison with three types of baselines, including single-agent execution, spatial communication and temporal communication. The best results are highlighted in bold, and the runner-ups are underlined.  {All methods, except for the single-agent category, utilize \textbf{five} \llmname{gpt-4}-based agents.}}
\vspace{-0.5em}
\label{tab:rq1_performance}
\renewcommand\tabcolsep{2.9pt}
\renewcommand\arraystretch{1.1}
  
\resizebox{\linewidth}{!}{
\begin{tabular}{l|cc|ccccccc}
\Xhline{1.2pt}
\rowcolor{CadetBlue!20} 
{\textbf{Method}} & \textbf{Spa.} & \textbf{Tem.} & \textbf{MMLU} & \textbf{GSM8K} & \textbf{MultiArith} & \textbf{SVAMP} & \textbf{AQuA} & \textbf{HumanEval} & {\textbf{Avg.}} \\
\Xhline{1.2pt}
Vanilla & \textcolor{darksalmon}{\XSolidBrush} & \textcolor{darksalmon}{\XSolidBrush} & 82.14 & 85.40 & 93.15 & 87.18 & 70.34 & 71.68 & 81.65\\
\hline

\rowcolor{gray!10}CoT  & \textcolor{darksalmon}{\XSolidBrush} & \textcolor{darksalmon}{\XSolidBrush} & 82.65\red{0.51} & 87.17\red{1.77} & 94.79\red{1.64} & 88.32\red{1.14} & 73.91\red{3.57} & 75.52\red{3.84} & 83.73\\

ComplexCoT  & \textcolor{darksalmon}{\XSolidBrush} & \textcolor{darksalmon}{\XSolidBrush} & 83.78\red{1.64} & 87.62\red{2.22} & 95.86\red{2.71} & 90.17\red{2.99} & 77.58\red{7.24} & 74.94\red{3.26} & 84.99\\

\rowcolor{gray!10}SC (CoT)  & \textcolor{darksalmon}{\XSolidBrush} & \textcolor{darksalmon}{\XSolidBrush} & 82.66\red{0.52} & 87.93\red{2.53} & 96.88\red{3.73} & 88.69\red{1.51} & 75.08\red{4.74} & 77.30\red{5.62} & 84.67 \\

SC (ComplexCoT)  & \textcolor{darksalmon}{\XSolidBrush} & \textcolor{darksalmon}{\XSolidBrush} & 83.65\red{1.51} & 86.14\blue{0.74} & 96.94\red{3.79} & 89.72\red{2.54} & 77.69\red{7.35} & 77.94\red{6.26} & 85.35\\

\hline

\rowcolor{gray!10}Chain & \textcolor{green(pigment)}{\Checkmark} & \textcolor{darksalmon}{\XSolidBrush} & 82.35\red{0.21} & 85.57\red{0.17} & 94.38\red{1.23} & 83.41\blue{3.77} & 70.94\red{0.60} & 80.88\red{9.20} & 92.92\\

Star & \textcolor{green(pigment)}{\Checkmark} & \textcolor{darksalmon}{\XSolidBrush} & 80.79\blue{1.35} & 85.55\red{0.15} & 93.79\blue{0.64} & 88.09\red{0.91} & 68.57\blue{1.77} & 75.65\blue{3.97} & 82.07\\

\rowcolor{gray!10}Tree & \textcolor{green(pigment)}{\Checkmark} & \textcolor{darksalmon}{\XSolidBrush} & 81.89\blue{0.25} & 84.56\blue{0.84} & 94.60\red{1.45} & 89.25\red{2.07} & 72.84\red{2.50} & 77.38\red{5.70}& 83.42 \\

Complete Graph & \textcolor{green(pigment)}{\Checkmark} & \textcolor{darksalmon}{\XSolidBrush} & 83.15\red{1.01} & 86.49\red{1.09} & 97.20\red{4.05} & 89.48\red{2.30} & 79.21\red{8.87} & 83.75\red{12.07} & 86.55\\

\rowcolor{gray!10}Layered Graph & \textcolor{green(pigment)}{\Checkmark} & \textcolor{darksalmon}{\XSolidBrush} & 78.41\blue{3.73} & 85.34\blue{0.06} & 95.04\red{1.89} & 88.61\red{1.43} & 73.18\red{2.84} & 80.38\red{8.70} &83.49\\

Random Graph & \textcolor{green(pigment)}{\Checkmark} & \textcolor{darksalmon}{\XSolidBrush} & 83.76\red{1.62} & 86.14\red{0.74} & 95.46\red{2.31} & 85.41\blue{1.77} & 74.07\red{3.73} & 82.66\red{10.98}& 84.58 \\

\rowcolor{gray!10}LLM-Blender & \textcolor{green(pigment)}{\Checkmark} &  \textcolor{darksalmon}{\XSolidBrush} & 81.22\blue{0.92} & 89.17\red{3.77} & 94.27\red{1.12} & 88.77\red{1.59} & 77.05\red{6.71} & - & 86.10 \\

GPTSwarm & \textcolor{green(pigment)}{\Checkmark} & \textcolor{darksalmon}{\XSolidBrush} & \underline{83.98}\red{1.84} & {89.74}\red{4.34} & \textbf{97.84}\red{4.69} & 86.42\blue{0.76} & 78.16\red{7.82} & 88.49\red{16.81} & 86.77 \\ 
\hline

\rowcolor{gray!10}LLM-Debate & \textcolor{darksalmon}{\XSolidBrush} & \textcolor{green(pigment)}{\Checkmark} & 83.69\red{1.55} & 90.23\red{4.83} & 96.27\red{3.12} & 90.56\red{3.38} & 77.52\red{7.18} & 83.79\red{12.11} & 87.01 \\

PHP & \textcolor{darksalmon}{\XSolidBrush} & \textcolor{green(pigment)}{\Checkmark} & 83.45\red{1.31} & 92.45\red{7.05} & 96.41\red{3.26} & 90.62\red{3.44} & 76.25\red{5.91} & 82.96\red{11.28} & 87.02\\

\rowcolor{gray!10}DyLAN & \textcolor{darksalmon}{\XSolidBrush} & \textcolor{green(pigment)}{\Checkmark} & 80.16\blue{1.98} & 88.16\red{2.76} & 94.27\red{1.12} & 87.40\red{0.22} & 74.16\red{3.82} & \underline{89.70}\red{18.02} & 84.48 \\

\hline

\ourmethod-C & \textcolor{green(pigment)}{\Checkmark} & \textcolor{green(pigment)}{\Checkmark} & \textbf{84.72}\red{2.58} & \underline{95.62}\red{10.22} & \underline{97.25}\red{4.10} & \textbf{91.85}\red{4.67} & \textbf{79.47}\red{9.13} & {89.38}\red{15.70} & \textbf{89.72}\\

\rowcolor{gray!10}\ourmethod-L & \textcolor{green(pigment)}{\Checkmark} & \textcolor{green(pigment)}{\Checkmark} & 83.50\red{1.36} & 93.78\red{8.38} & 96.39\red{3.24} & 89.58\red{2.40} & 78.44\red{8.10} & 88.61\red{16.93} & 88.38 \\

\ourmethod-R & \textcolor{green(pigment)}{\Checkmark} & \textcolor{green(pigment)}{\Checkmark} & 83.94\red{1.80} & \textbf{95.83}\red{10.43} & 96.30\red{3.15} & \underline{91.68}\red{4.50} & \underline{78.60}\red{8.26} & \textbf{90.30}\red{18.62} & \underline{89.44}\\

\Xhline{1.2pt}
\end{tabular}
}
\vspace{-1.5em}
\end{table*}

\vspace{-0.4em}
\subsection{Experimental Setup}\label{sec:exp_setup}
\vspace{-0.4em}
\paragraph{Tasks and Benchmarks} In our experiments, we test the performance of \ourmethod on three types of reasoning tasks and the corresponding logically challenging benchmarks: \textbf{{(1) General Reasoning}}: We opt for MMLU~\citep{mmlu} dataset; \textbf{{(2) Mathematical Reasoning}}: We select GSM8K~\citep{arXiv2021_Verifier-Math}, MultiArith~\citep{roy2016solving}, SVAMP~\citep{patel2021nlp} and AQuA~\citep{ling2017program} to verify the mathematical reasoning capacity; \textbf{{(3) Code Generation}}: We use the HumanEval~\citep{human-eval} to test the function-level code generation ability.
\vspace{-0.8em}
\paragraph{Baselines} We compare \ourmethod with three series of multi-agent communication paradigms, namely: \textbf{(1) Single agent execution methods}, including Chain-of-Thought prompting (CoT; \cite{cot}), (2) Complexity-based prompting (ComplexCoT; \cite{fu2022complexity}), and (3) Self-Consistency (SC; \cite{wang2023selfconsistency}); \textbf{(2) Spatial communication methods}, including chain, tree, star, complete graph, layered graph and random graph\footnote{Detailed explanations of these topologies are placed in \Cref{app:topology}.} from MacNet~\citep{qian2024scaling}, LLM-Blender~\citep{blender},  and GPTSwarm~\citep{zhuge2024gptswarm}; \textbf{(3) Temporal communication methods}, including PHP~\citep{PHPrompting}, LLM-Debate~\citep{arXiv2023_MultiAgent-Debate}, DyLAN~\citep{arXiv2023_Dynamic-LLM-Agent}. Detailed introductions and implementations of the baselines are in \Cref{app:exp_baseline}.

\vspace{-1.em}

\paragraph{Implementation Details} We accessed the GPT models via the OpenAI API, and mainly tested \llmname{gpt-3.5-turbo-0301} (\llmname{gpt-3.5}) and \llmname{gpt-4-1106-preview} (\llmname{gpt-4}) with different communication topologies. We set the temperature at 1 during the generation. We set the dialogue round $K=2$ for mathematical and general reasoning tasks, and $K=4$ for code generation tasks. For multi-query settings, we vary $Q'\in\{5,10\}$. We generate different agent profiles using \llmname{gpt-4} for individual agents. More experimental details are in \Cref{app:exp}.

\vspace{-0.6em}
\subsection{Performance \& Cost Comparison (RQ1)}\label{exp:perf_cost}

\begin{figure*}[!t]
\centering
\includegraphics[width=\linewidth]{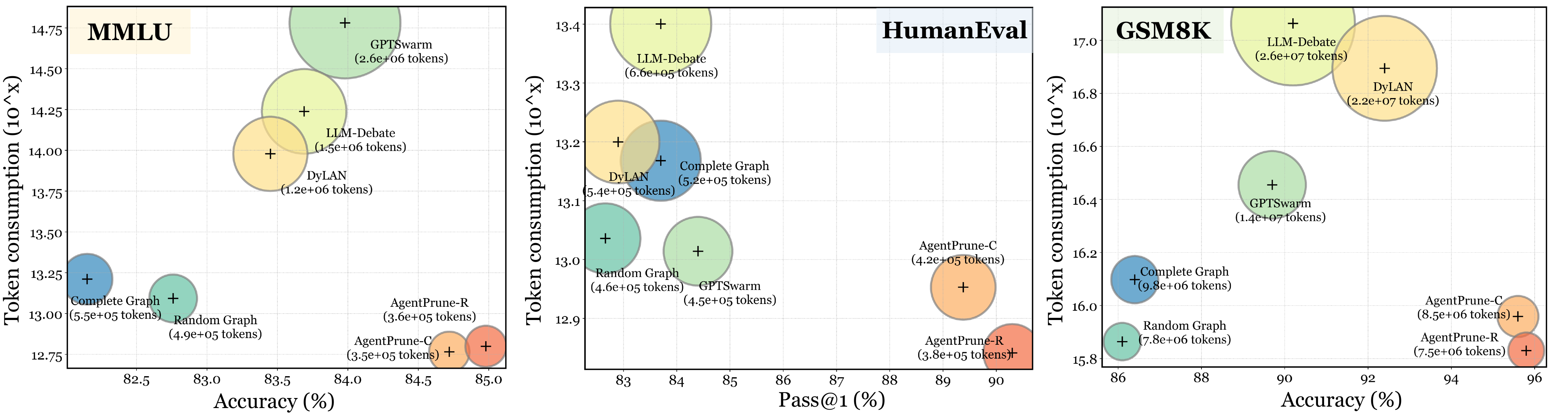}
\vspace{-1.5em}
\caption{\textbf{Visualization of performance and prompt token consumption.} This scatter plot illustrates the performance metrics and \textbf{prompt token consumption} of different multi-agent communication topologies across MMLU, HumanEval, and GSM8K. The diameter of each point is proportional to its y-axis value.}
\label{fig:scatter_1}
\vspace{-1.5em}
\end{figure*}

\begin{wraptable}{r}{0.45\textwidth} \vspace{-1.2em}
 \centering
  \centering
  \caption{
  \textbf{Performance on the HumanEval Benckmark} with more advanced baselines.
  }
  \label{tab:code}
  \vspace{-0.8em}
  \renewcommand\tabcolsep{6.5pt}
  \renewcommand\arraystretch{1.1}
  \footnotesize 
  \begin{tabular}{l|cc} 
    \Xhline{1.2pt}
    \rowcolor{CadetBlue!20} 
    \textbf{Method} & \textbf{Pass@1} & \textbf{$\Delta$} \\
    \Xhline{1pt}
   Vanilla & 71.68 & - \\
    \rowcolor{gray!10}AutoGen~[\citeyear{autogen}] & 85.41 & {\color{RedOrange} $\uparrow$11.97} \\
    Reflexion~[\citeyear{reflexion}] & 91.40 & {\color{RedOrange} $\uparrow$19.72} \\
   \rowcolor{gray!10}CodeT+Parsel~[\citeyear{zelikman2023parsel}] &  85.10 & {\color{RedOrange} $\uparrow$13.42} \\
   MetaGPT~[\citeyear{meta-gpt}] & 85.90 & {\color{RedOrange} $\uparrow$14.22}\\
    \rowcolor{gray!10}ANPL~[\citeyear{huang2024anpl}] & 86.60 & {\color{RedOrange} $\uparrow$14.92}\\
    \hline
    \ourmethod-C  & {89.38} & {\color{RedOrange} $\uparrow$17.70}\\
   \rowcolor{gray!10}\ourmethod-R  & {90.30} & {\color{RedOrange} $\uparrow$18.62}\\
    \Xhline{1.2pt}
  \end{tabular}
  \vspace{-0.3em}
\end{wraptable}
\vspace{-0.5em}
To evaluate whether \ourmethod achieves a dual benefit of token savings and task completion, we integrate it with three predefined spatial communication topologies: the complete graph, layered graph, and random graph, denoted as \ourmethod-C, \ourmethod-L, and \ourmethod-R, respectively. 
For the temporal communication topology, we consistently employ the fully connected LLM-Debate-style structure. 
\Cref{tab:rq1_performance,tab:code} presents a performance comparison of various communication paradigms within \textit{five} \llmname{gpt-4}-based multi-agent systems, and \Cref{fig:scatter_1,fig:scatter_2,fig:scatter_3,fig:scatter_4} visualizes the performance and  token cost of different methods. Our observations (Obs.) are as follows: 
\textbf{Obs.}\ding{182} \textbf{Not all multi-agent topologies consistently deliver collective intelligence.} As illustrated in \Cref{tab:rq1_performance}, certain topologies, such as star/tree structures, fail to consistently improve performance for multi-agent systems, even resulting in performance drops of $0.17\%\sim3.97\%$. In contrast, single-agent prompting methods like CoT or ComplexCoT demonstrate much more stable and significant improvements.
\textbf{Obs.}\ding{183} \textbf{The high performance of existing multi-agent systems comes at a substantial economical cost.}
From \Cref{tab:rq1_performance}, we observe that the top-performing baselines, GPTSwarm and DyLAN, achieve \textit{pass@1} improvements of $16.81\%$ and $18.02\%$ on HumanEval, respectively; however, this is accompanied by extremely high economic costs. As shown in \Cref{fig:scatter_1}, the prompt token consumption of GPTSwarm and DyLAN is 
$2.4\sim5.3\times$ that of the random graph structure. \textbf{Obs.}\ding{184} \textbf{\ourmethod achieves a double win in economic savings and utility.} Among the three variants, \ourmethod-R delivers consistently impressive performance, achieving $90.3\%$ on HumanEval and $95.8\%$ on GSM8K. Importantly, this performance does not come at a high token cost: on both HumanEval and GSM8K, the token consumption of \ourmethod is less than $40\%$ that of DyLAN. Overall, \ourmethod excels in both task completion and token efficiency.
\vspace{-0.6em}
\subsection{Plug-in into Existing Frameworks (RQ2)}\label{sec:plugin}
\vspace{-0.6em}
As a plug-in, \ourmethod can be seamlessly combined with mainstream multi-agent pipelines, effectively reducing the economic costs associated with LLM token throughput while maintaining the original performance levels. To validate our argument, we combined \ourmethod with two representative LLM-MA frameworks, AutoGen and GPTSwarm. 
With the results presented in \Cref{tab:combine_cost_5} and \Cref{tab:combine_cost_3}, we offer the following two key observations: \textbf{Obs.}\ding{185} \textbf{Scaling multi-agent collaboration is costly.} Comparing \Cref{tab:combine_cost_5} and \Cref{tab:combine_cost_3}, we observe that for the GPTSwarm on the GSM8K dataset, optimizing a three-agent system incurs a cost of $\$97.23$, while the expense for a five-agent system skyrockets to $\$234.76$, with the total token count reaching $1.7e+7$. AutoGen, on the other hand, has relatively lower costs because it does not involve the iterative optimization of the communication topology as extensively as GPTSwarm~\citep{zhuge2024gptswarm}. Nevertheless, it still requires \$73.21 on the GSM8K benchmark, which comprises up to 8.5K data entries.
\textbf{Obs.}\ding{186} \textbf{\ourmethod is an economically friendly assistant.} When applied to HumanEval+AutoGen, \ourmethod achieves a $36\%$ reduction in prompt tokens and saves $\$1.486$. In tasks with larger datasets, the economic savings become even more pronounced: on GSM8K+GPTSwarm, \ourmethod reduces $60.6\%$ of the prompt token consumptions and saves a cost of up to $\$177.58$, with even a performance increase of $0.84\%$. Overall, \ourmethod serves as a token-efficient plug-in, effectively fostering the development of larger and more cost-effective multi-agent systems.

\begin{table}[t!]
  \centering
  \caption{
  \textbf{Performance and cost comparison before/after combining \ourmethod.} We evaluated the performance and economical cost of \ourmethod in conjunction with two classic multi-agent systems, under a \textbf{five} \llmname{gpt-4}-based setting. ``\# Prompt tokens'' refers to the total number of tokens input, while ``\# Completion tokens'' accounts for the total number of tokens output by the API.
  }
  \label{tab:combine_cost_5}
  \vspace{-0.5em}
  \renewcommand\tabcolsep{5.5pt}
  \renewcommand\arraystretch{1.1}
  \footnotesize 
  \resizebox{\linewidth}{!}{
  \begin{tabular}{lc|cccc} 
    \Xhline{1.2pt}
    \rowcolor{CadetBlue!20} 
    \textbf{Dataset} & \textbf{Method} & \textbf{Performance} & \textbf{\# Prompt Tokens}  & \textbf{\# Completion Tokens} & \textbf{Cost (USD)}\\
    \Xhline{1pt}
    \multirow{2}{*}{MMLU} & AutoGen & $82.13$ & $486,034$ & $89,224$ & $\$7.537$ \\
    & +\ourmethod & $82.78({\color{RedOrange} \uparrow 0.65})$ & $349,583(\color{RedOrange} 71.9\%)$  & $86,582$ & $\$6.093$\\
    \hline
     \rowcolor{gray!10}& AutoGen & $85.41$ & $492,273$ & $130,196$ & $\$8.828$ \\
   \rowcolor{gray!10} \multirow{-2}{*}{HumanEval}& +\ourmethod & $86.65({\color{RedOrange} \uparrow 1.24})$ & $315,105({\color{orange} 64.0\%})$ & $139,714$ & $\$7.342$\\
    \hline
     & AutoGen & $90.06$ & $4,327,740$ & $998,042$ & $\$73.21$  \\
    \multirow{-2}{*}{GSM8K}& +\ourmethod & $92.85({\color{RedOrange} \uparrow 2.79})$ & $3,791,251 ({\color{orange} 59.9\%})$ & $1,156,884$ & $\$59.60$\\
    \hline
    \rowcolor{gray!10} & GPTSwarm & $83.98$ & $3,055,230$ & $569,124$ & $\$47.60$ \\
    \rowcolor{gray!10}\multirow{-2}{*}{MMLU}& +\ourmethod & $83.05({\color{BlueGreen} \downarrow 0.93})$ & $990,312({\color{orange} 32.4\%})$ & $439,551$ & $\$23.05$\\
    \hline
    
     & GPTSwarm & $84.49$ & $2,736,136$ & $1,004,616$ & $\$57.49$\\
    \multirow{-2}{*}{HumanEval}& +\ourmethod & $84.96({\color{RedOrange} \uparrow 0.47})$ & $745,617({\color{orange} 27.2\%})$ & $745,926$ & $\$29.80$\\
    \hline
    \rowcolor{gray!10} & GPTSwarm & $89.74$ & $14,005,945$ & $3,156,916$ & $\$234.76$ \\
    \rowcolor{gray!10}\multirow{-2}{*}{GSM8K}& +\ourmethod & $90.58({\color{RedOrange} \uparrow 0.84})$ & $3,526,035({\color{orange} 39.4\%})$ & $730,552$ & $\$57.17$\\
    \hline
    \Xhline{1.2pt}
  \end{tabular}}
\end{table}

\begin{figure*}[!t]
\vspace{-0.6em}
\centering
\includegraphics[width=\linewidth]{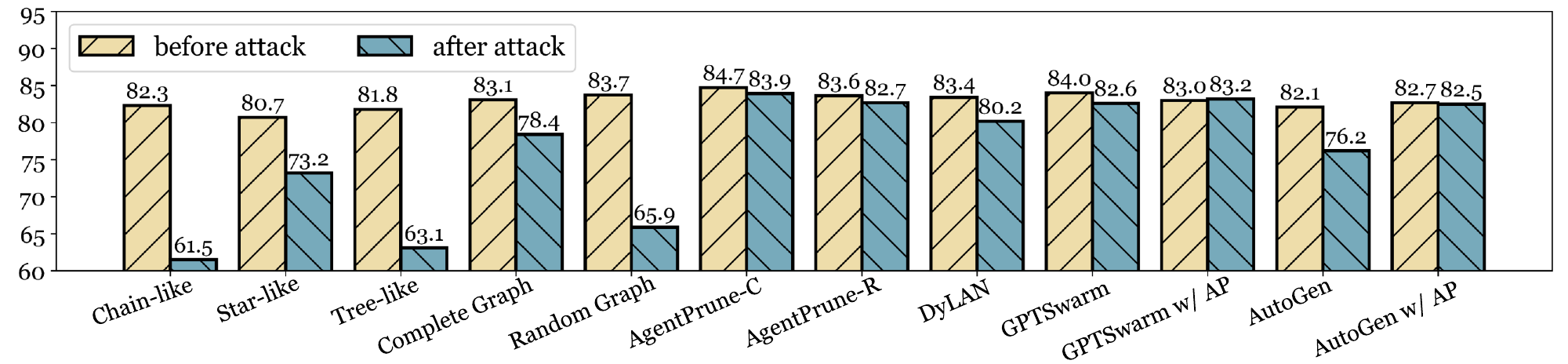}
\vspace{-1.9em}
\caption{\textbf{Performance under adversarial attack.} We compare the accuracy (\%) of various multi-agent frameworks before and after \textit{prompt attacks} on MMLU. ``w/ AP'' indicates the integration with \ourmethod.}
\label{fig:attack}
\vspace{-1.7em}
\end{figure*}

\vspace{-0.6em}
\subsection{Robustness Verification (RQ3)}\label{exp:robust}
\vspace{-0.6em}
\ourmethod can not only eliminate \textit{unnecessary} communications but also remove \textit{malicious} messages. To validate this, we design two types of adversarial attacks for the multi-agent frameworks: the \textbf{agent prompt attack} and \textbf{agent replacement attack}. The former attacks the role prompts of the agents, while the latter attacks the LLM's generation process, with detailed implementation elaborated in \Cref{app:attack}. We observe that: \textbf{Obs.}\ding{187} \textbf{Existing LLM-MA frameworks often lack adversarial robustness.} Despite variations in performance, most frameworks experience significant declines when subjected to both types of attacks. As shown in \Cref{fig:attack,fig:attack_2}, the chain-like structure suffers a performance drop of up to $20.8\%$ due to its oversimplistic topology. AutoGen and DyLAN similarly experience accuracy declines ranging from $3.2\%$ to $6.2\%$. \textbf{Obs.}\ding{188} \textbf{\ourmethod significantly enhances multi-agent robustness.} \Cref{fig:attack} demonstrates that combining \ourmethod with a complete graph not only improves performance ($83.1\%\rightarrow84.7\%$), but also increases robustness under agent prompt attacks ($78.4\%\rightarrow 83.9\%$). Additionally, \ourmethod successfully boosts the robustness of DyLAN and AutoGen by up to $6.3\%$. The impact of \ourmethod on GPTSwarm is relatively marginal, due to its inherent defenses against adversarial agents. Overall, \ourmethod serves as an easy-to-use enhancer for multi-agent robustness.

\vspace{-0.8em}
\subsection{Experimental Analysis}\label{sec:exp_others}
\vspace{-0.6em}
\paragraph{Ablation Study} We ablate agent profiling and low-rank regularization in \ourmethod, with details presented in \Cref{tab:ablation} and \Cref{app:ablation}. Our key finding is that (1) the utility of agent profiling varies across different datasets, demonstrating a more pronounced effect on general reasoning and code generation tasks, while being relatively less significant in math reasoning; (2) low-rank sparsity consistently facilitates the optimization of the communication topology.
\vspace{-1.2em}
\paragraph{Sensitivity Analysis \textit{and} Case Study} We present the parameter sensitivity analysis concerning two hyperparameters in \Cref{app:sensitivity}, and provide extensive visualizations on \ourmethod's pruning process and optimized communication structure in \Cref{app:case_study}.

\vspace{-0.8em}
\section{Related Work}
\vspace{-0.5em}
\paragraph{LLM-agent Collaboration} Collaboration between multiple LLM-based agents has emerged as a promising approach to enhance the capabilities of individual LLMs~\citep{arXiv2023_MultiAgent-Debate, arXiv2023_MultiAgent-Debate_2, multi-persona}. As stated in \Cref{sec:intro}, current multi-agent communication methods can be categorized into two types: \ding{182} \textbf{Intra-dialogue (spatial) communication} focuses on how different agents exchange messages within a single dialogue round. Common structures include (1) \textit{Direct output}, where functioning agents do not communicate with each other, adopted by systems like LATM~\citep{zhang2023astools}, LLM-Debate~\citep{arXiv2023_MultiAgent-Debate}; (2) \textit{Chain}, employed by ChatDev~\citep{software-dev}, MetaGPT~\citep{meta-gpt} and L2MAC~\citep{holt2024l2mac}; (3) \textit{Tree}, where an administrative agent (usually refered to as commander, manager, \textit{etc}.) controls subordinate agents, adopted by AutoGen~\citep{autogen}, SecurityBot~\citep{yan2024depending}, and MiniGrid~\citep{zhou2023large}; and (4) \textit{Graph,} employed by LLM-Blender~\citep{blender}, ChatEval~\citep{chateval}, MacNet~\citep{qian2024scaling} and GPTSwarm~\citep{zhuge2024gptswarm}; \ding{183} \textbf{Inter-dialogue (temporal) communication} focuses on how information is passed between different rounds of utterances. Common topologies include (1) \textit{Full transmission}, where every agent receives the utterances of all agents from the previous round, as used by LLM-Debate~\citep{arXiv2023_MultiAgent-Debate}; (2) \textit{Partial transmission}, where some responses are filtered through scoring or rating mechanisms, adopted by PHP~\citep{PHPrompting} and DyLAN~\citep{arXiv2023_Dynamic-LLM-Agent}; (3) \textit{Summarization}, where dialogue history is compressed and summarized for the next round of communication, as seen in Reflexion~\citep{reflexion}, ICL-AIF~\citep{bargaining-feedback}, AgentVerse~\citep{chen2023agentverse}, CoMM~\citep{chen2024comm}, Corex~\citep{sun2023corex}, and MAD~\citep{arXiv2023_MultiAgent-Debate_2}.
\vspace{-1em}
\paragraph{Agents as Graphs} Learning to facilitate communication via learning graph connectivity is a long-standing and viable approach to enhance multi-agent cooperation~\citep{pesce2023learning,hu2024magraph}. In the pre-LLM era, numerous efforts explored optimal communication graph structures for reinforcement learning-based multi-agents with graph diffusion~\citep{pesce2023learning}, weighted GNN~\citep{liu2022temporal}, or transformers~\citep{hu2024magraph}. In the emerging wave of LLM-powered agents, attempts that leverage graphs for modeling agent-agent interaction also exist: ChatEval~\citep{chateval} and AutoGen~\citep{autogen} implicitly adopt graph structures to describe "simultaneous talk", and STOP~\citep{zelikman2023self} and DSPy~\citep{khattab2023dspy} optimize both the prompts and the inference structure together. MacNet~\citep{qian2024scaling} and GPTSwarm~\citep{zhuge2024gptswarm} model agent communication via directed acyclic graphs (DAG). However, none of these approaches simultaneously optimize both intra- and inter-dialogue communication structures, and they often result in even increased token consumption.

\vspace{-0.7em}
\section{Conclusion}
\vspace{-0.9em}
This paper makes the first attempt towards a high-performance and token-efficient LLM-powered multi-agent system. We propose an economical, simple, and robust multi-agent communication pipeline, termed \ourmethod, which can be harmoniously embedded into mainstream multi-agent frameworks while effectively pruning the \textit{communication redundancy} that we have identified and defined. \ourmethod achieves performance comparable to, or even superior to, the original systems with significantly smaller token throughput and economic costs. We believe that \ourmethod can facilitate the advancement toward larger-scale collective intelligence.

\bibliography{iclr2025_conference}
\bibliographystyle{iclr2025_conference}

\appendix

\section{DAG Sampling Function}

\begin{algorithm}[H]
\SetKwFunction{Facyclic}{is\_acyclic}
\SetKwFunction{Fcantopo}{can\_topo\_sort}
\SetKwFunction{Ffindcycle}{find\_cycle}
\SetKwFunction{Frandom}{random\_choice}
\SetKwFunction{Fremove}{remove\_edge}

\caption{Sample DAG from spatial communication graph}\label{algo:dag}
\KwIn{Spatial communication graph $\mathcal{G}^\mathcal{S}=\{\mathcal{V},\mathcal{E}^\mathcal{S}\}$}
\KwOut{A directed acyclic graph $\hat{\mathcal{G}}^\mathcal{S}$}

$\hat{\mathcal{G}} \gets \mathcal{G}^\mathcal{S}$ \tcp{{Create a copy of $\mathcal{G}^\mathcal{S}$}}

\While{\rm{not} \Facyclic($\hat{\mathcal{G}}$)}{
    \tcc{Use DFS to locate cycle}
    
    cycle $\gets$ \Ffindcycle($\hat{\mathcal{G}}$) 
    
    e $\gets$ \Frandom(cycle) 
    \tcp{Randomly select an edge from the cycle}
    $\hat{\mathcal{G}} \gets \hat{\mathcal{G}}.$\Fremove(e)
    \tcp{Remove the selected edge from $G'$}
}
\Return $\hat{\mathcal{G}}^\mathcal{S} \gets \hat{\mathcal{G}}$
\end{algorithm}

\section{Notations}\label{app:notation}

We conclude the commonly used notations in \Cref{tab:notation} for reference.

\begin{table}[!h]\footnotesize
  \centering
  \caption{The notations that are commonly used throughout the manuscript.}
   \setlength{\tabcolsep}{16pt} 
   \renewcommand\arraystretch{1.15} %
  \vspace{-2mm}
    \begin{tabular}{cc}
        \Xhline{1.2pt}
    \rowcolor{CadetBlue!20} 
    \textbf{Notation} & \textbf{Definition} \\
     \Xhline{1.pt}
    $\mathcal{G}=(\mathcal{V},\mathcal{E})=\{\mathcal{G}^\mathcal{S},\mathcal{G}^\mathcal{T}\}$          & the spatial-temporal communication graph\\
    $\mathcal{V}=\{v_1,v_2,\cdots,v_{|\mathcal{V}|}\}$ & the set of nodes (agents)\\
    $\mathcal{E} = \mathcal{E}^{\mathcal{S}}\cup \mathcal{E}^{\mathcal{T}}$ &  the overall edge set\\
    $\mathcal{E}^{\mathcal{S}}\subseteq\mathcal{V}^{(t)}\times\mathcal{V}^{(t)}$ & the spatial edge set\\
    $\mathcal{E}^{\mathcal{T}}\subseteq \mathcal{V}^{(t-1)}\times\mathcal{V}^{(t)}$ & the temporal edge set\\
    $\texttt{Base}_i$ & the LLM base utilized by agent $v_i$\\
    $\texttt{Role}_i$ & the predefined responsibilities or roles of agent $v_i$\\
    $\texttt{State}_i$ & the state of agent $v_i$\\
    $\texttt{Plugins}_i=\{\texttt{F}_j,\texttt{C}_j\}_{j=1}^{P}$ & the plugins available to agent $v_i$\\
    $e^{\mathcal{S}}_{ij} = (\mathbf{M}_{ij}, \mathbf{O}_{ij})$ & the spatial edge from $v_i$ to $v_j$\\

$e^{\mathcal{T}}_{ij}$ & the temporal edge from $v_i$ to $v_j$\\

$\mathcal{N}^\mathcal{T}(v_i) = \{v_j\;|\;(j,i)\in\mathcal{E}^\mathcal{T}\}$& the temporal (in-)neighbors of $v_i$\\

$\mathcal{N}^\mathcal{S}(v_i) = \{v_j\;|\;(j,i)\in\mathcal{E}^\mathcal{S}\}$ & the spatial (in-)neighbors of $v_i$\\

$\mathbf{M}^{(t)}_i$ & the rationale or answers provided by $v_i$ at the $t$-th epoch\\

$\mathcal{G}^\text{sub} = (\mathcal{V}, \mathcal{E}' \cup \mathcal{E}'')$ & the sparsified communication topology\\

$\mathbf{S}^\mathcal{S}, \mathbf{S}^\mathcal{T}\in\mathbb{R}^{|\mathcal{V}|\times|\mathcal{V}|}$ & the spatial and temporal graph masks\\

$\mathbf{A}^{\mathcal{S}}\in \{0,1\}^{|\mathcal{V}|\times|\mathcal{V}|}$ & the predefined spatial communication topology\\

$\mathbf{A}^{\mathcal{T}}\in \{0,1\}^{|\mathcal{V}|\times|\mathcal{V}|}$ & the predefined temporal communication topology\\

$\tilde{\mathcal{G}}^\mathcal{S}$ & the parameterized spatial graph\\

$\hat{\mathcal{G}}^\mathcal{S}$ & the parameterized spatial graph after DAG sampling\\

$\tilde{\mathcal{G}}^\mathcal{T}$ & the parameterized temporal graph\\

$\phi(\cdot)$ & the utility evaluation function\\

$\mathbf{B}^\mathcal{S}\in \{0,1\}^{|\mathcal{V}|\times|\mathcal{V}|}$ &  the obtained binary spatial mask\\

$\mathbf{B}^\mathcal{T}\in \{0,1\}^{|\mathcal{V}|\times|\mathcal{V}|}$ & the obtained binary temporal mask\\

$K$ & the total number of dialogue rounds\\

$K'$ & the dialogue round after which pruning takes place\\

$Q$ & the total number of queries\\

$Q'$ & the number of queries after which pruning takes place\\
        \Xhline{1.2pt}

    \end{tabular}%
  \label{tab:notation}%
\end{table}%

\section{Algorithm Workflow}\label{app:workflow}
We conclude the overall algorithm workflow of \ourmethod in \cref{algo:agentprune_1}.

\begin{algorithm}[!ht]
\DontPrintSemicolon
\SetAlgoLined
\LinesNumbered
\SetKwFunction{Fstop}{MeetEndCondition()}
\SetKwFunction{Ftopo}{TopologicalSort}
\SetKwFunction{Faggregate}{AggregateSolution}
\KwIn{Query $q$, Communication graph $\mathcal{G}=\{\mathcal{G}^\mathcal{S},\mathcal{G}^\mathcal{T}\}$, Maximum rounds of iterations $K$, Rounds for optimization $K'$, Initial masks $\mathbf{S}^\mathcal{S},\mathbf{S}^\mathcal{T}$}

$\mathbf{A}(\{\Tilde{\mathcal{G}}^\mathcal{S},\Tilde{\mathcal{G}}^\mathcal{T}\}) \leftarrow \{\mathbf{A}^\mathcal{S}\odot \mathbf{S}^\mathcal{S}, \mathbf{A}^\mathcal{T}\odot \mathbf{S}^\mathcal{T}\}$

\tcc{Optimizing spatial-temporal communication topology}
\For{\rm{iteration} $t\leftarrow1$ \KwTo $K'$}{

$\hat{\mathcal{G}}^\mathcal{S} = (\mathcal{V},\mathcal{E}^\mathcal{S}\cup\mathcal{E}^\mathcal{T})\leftarrow \texttt{DAGSampling}(\Tilde{\mathcal{G}}^\mathcal{S})$

\For{$v_i$ \rm{in} $\Ftopo(\mathcal{V})$}{

Obtain temporal (in-)neighbors $\mathcal{N}^\mathcal{T}(v_i) \leftarrow \{v_j\;|\;(j,i)\in\mathcal{E}^\mathcal{T}\}$

Obtain spatial (in-)neighbors $\mathcal{N}^\mathcal{S}(v_i) \leftarrow \{v_j\;|\;(j,i)\in\mathcal{E}^\mathcal{S}\}$

$\boldsymbol{m}^\mathcal{T}\gets \{\mathbf{M}_j\;|\;v_j\in\mathcal{N}^\mathcal{T}(v_i)\}, \boldsymbol{m}^\mathcal{S} \gets \{\mathbf{M}_{ij}\;|\;v_j\in\mathcal{N}^\mathcal{S}(v_i)\}$ 

$\mathbf{M}_i^{(t)} \sim \mathcal{P}_{\theta} (\mathbf{M}_i\;|\;q, \texttt{Role}^{(t)}_i,\texttt{State}^{(t)}_i, \boldsymbol{m}^\mathcal{T}, \boldsymbol{m}^\mathcal{S})$ 
}
$a^{(t)} \gets \Faggregate(\mathbf{M}_1^{(t)}, \mathbf{M}_2^{(t)},\cdots, \mathbf{M}_{|\mathcal{V}|}^{(t)})$ 

Update $\mathbf{S}^\mathcal{S}, \mathbf{S}^\mathcal{T}$ according to \Cref{eq:objective}

}

\tcc{One-shot pruning spatial-temporal communication topology}
$\mathbf{B}^\mathcal{S} = \mathbb{I}(\mathbf{A}^\mathcal{S} \neq 0 \wedge \operatorname{TopK}(\mathbf{S}^\mathcal{S}, |\mathbf{A}^\mathcal{S}|\times \left(1-p\%\right)))$

$\mathbf{B}^\mathcal{T} = \mathbb{I}(\mathbf{A}^\mathcal{T} \neq 0 \wedge \operatorname{TopK}(\mathbf{S}^\mathcal{T}, |\mathbf{A}^\mathcal{T}|\times \left(1-p\%\right)))$

Obtain $\mathcal{G}^\text{sub}$, where $\mathbf{A}(\mathcal{G}^\text{sub}) 
= \{\mathbf{A}^\mathcal{S} \odot \mathbf{B}^\mathcal{S}, \mathbf{A}^\mathcal{T}\odot \mathbf{B}^\mathcal{T}\}$

\tcc{Fixing the topology for subsequent iterations}

\For{\rm{iteration} $t\leftarrow K'$ \KwTo $K$}{
Use $\mathcal{G}^\text{sub}$ for multi-agent dialogues as in \Cref{algo:llmma}
}

\Return $a^{(t)}$ as the final solution
\caption{Execution pipeline of LLM-MA systems combined with \ourmethod.}
\label{algo:agentprune_1}
\end{algorithm}

\section{Multi-Query Training of \ourmethod}\label{app:multi_query}

For complex tasks such as repository-level code generation~\citep{software-dev}, multi-turn dialogues ($K>5$) are often essential. In such cases, utilizing $K' \in \{1,2\}$ rounds to optimize the topology and subsequently continue the dialogue for $K-K'$ rounds is reasonable. However, for simpler tasks that involve numerous queries, such as multiple-choice answering~\citep{agashe2023evaluating} or basic mathematical problems~\citep{arXiv2021_Verifier-Math}, previous studies~\citep{yin2023exchange,qian2024scaling} suggest that typically only $1$ to $2$ dialogue rounds are needed. In this context, prior dialogue-level optimization is no longer applicable. To better adapt \ourmethod to such circumstances, we propose a query-level optimization paradigm for \ourmethod.

Given a benchmark consisting of $Q$ queries, any LLM-MA framework processes these $Q$ queries sequentially to provide solutions one by one. We utilize the initial $Q' (Q' << Q)$ queries as a "training phase," collaboratively optimizing the spatio-temporal communication topology while leveraging multiple agents for reasoning and evaluation. Following this, we perform one-shot pruning as described in \Cref{eq:oneshot}. The fixed topology $\mathcal{G}^\text{sub}$ is then employed for the reasoning and evaluation of the remaining $(Q - Q')$ queries. We also refer to this approach as \textbf{query-level optimization}, in contrast to the \textbf{dialogue-level optimization} discussed in the main text. The distinction between the two lies in their focus: the latter concentrates on resolving a single query by utilizing several initial utterances to derive the topology, while the former considers the entire benchmark, employing a few early queries to inform the topology.

\section{Cost Analysis}\label{app:cost_analysis}
In \Cref{sec:cost}, we present a token-saving analysis in a single-query setting. In this section, we provide a cost analysis for \ourmethod in a multi-query optimization context. Given a communication graph \( \mathcal{G} \) and a benchmark with \( Q \) queries, we assume that the LLM-MA framework iterates for \( K \) dialogue rounds for each query. Furthermore, we denote the average token count per spatial, temporal, and query message as \( c_\mathcal{S} \), \( c_\mathcal{T} \), and \( c_q \), respectively. Hence, the total token consumption of the vanilla system can be expressed as:
\begin{equation}
C_\mathcal{G} = QK\left[ c_\mathcal{S} |\mathcal{E}^\mathcal{S}| + c_\mathcal{T} |\mathcal{E}^\mathcal{T}| + C_q|\mathcal{V}| \right]    
\end{equation}
When utilizing \ourmethod, the LLM-MA framework processes the initial \( Q' \) queries while simultaneously optimizing the spatial-temporal connectivity. The token cost for this phase is:
\begin{equation}
MQ'K\left[ c_\mathcal{S} |\mathcal{E}^\mathcal{S}| + c_\mathcal{T} |\mathcal{E}^\mathcal{T}| + C_q|\mathcal{V}| \right].
\end{equation}
After pruning \( \mathcal{G}^\mathcal{S} \) and \( \mathcal{G}^\mathcal{T} \), we use the obtained \( \mathcal{G}^\text{sub} \) to solve the remaining \( Q - Q' \) queries, with a cost of:
\begin{equation}
(Q-Q')K\left[ (1-p\%)\cdot\left(c_\mathcal{S} |\mathcal{E}^\mathcal{S}|+c_\mathcal{T} |\mathcal{E}^\mathcal{T}|\right) + C_q|\mathcal{V}| \right].
\end{equation}
Overall, the token savings of \ourmethod in a multi-query setting can be expressed as:
\begin{equation}
\Delta = (p\%\cdot Q + (1-p\% - M)Q')K\left(c_\mathcal{S} |\mathcal{E}^\mathcal{S}|+c_\mathcal{T} |\mathcal{E}^\mathcal{T}|\right)+(1-M)Q'KC_q|\mathcal{V}|
\end{equation}

\section{Exisiting Spatial Communication Topologies}\label{app:topology}

In this section, we introduce several existing spatial communication topologies, including chain, tree, star, complete graph, layered graph, random graph, and LLM-Blender.

\subsection{Chain Structure}

\begin{figure*}[!h]
\centering
\includegraphics[width=0.4\linewidth]{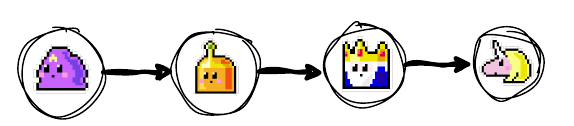}
\caption{Demonstration of \textbf{chain} structure as spatial communication topology.}
\label{fig:topo_chain}
\end{figure*}

The chain structure (in \Cref{fig:topo_chain}) is one of the most widely utilized communication architectures in contemporary multi-agent systems, as demonstrated by its application in ChatDev~\citep{software-dev}, MetaGPT~\citep{meta-gpt}, and L2MAC~\citep{holt2024l2mac}. In this architecture, the first agent receives input from the user, transforms it into new instruction, and subsequently forwards it to the next agent. For instance, in MetaGPT, user instructions are initially sent to the first agent, termed the "product manager," with information progressively relayed to subsequent agents, such as the architect agent, engineer agent, and QA engineer agent. Generally, the final agent in the chain provides a solution to the user's request.

\subsection{Tree Structure}

\begin{figure*}[!h]
\centering
\includegraphics[width=0.6\linewidth]{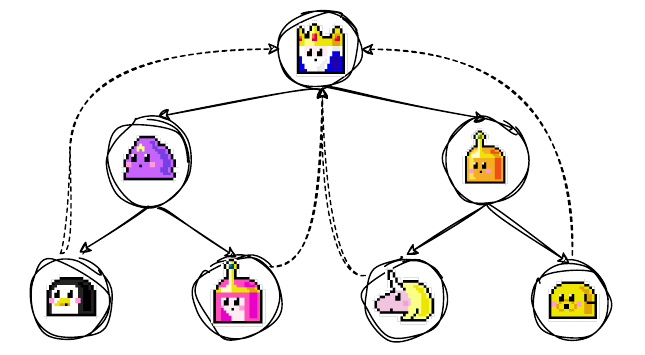}
\caption{Demonstration of \textbf{tree} structure as spatial communication topology.}
\label{fig:topo_tree}
\end{figure*}

In a tree-like multi-agent pipeline, as shown in \Cref{fig:topo_tree}, an administrative agent (commonly referred to as a teacher, commander, manager, etc.) oversees subordinate agents, which typically have distinct responsibilities. Ultimately, these subordinate agents submit their outputs to the administrative agent for final evaluation. Notable works employing this structure include AutoGen~\citep{autogen}, SecurityBot~\citep{yan2024depending}, and MiniGrid~\citep{zhou2023large}. For instance, in AutoGen (A4: Multi-Agent Coding), there exists a Commander agent alongside a Safeguard agent. The Writer is responsible for crafting the code and its interpretation, the Safeguard ensures safety (e.g., preventing information leaks and avoiding malicious code), and the Commander executes the code.

\subsection{Star Structure}\label{app:struc_star}
\begin{figure*}[!h]
\centering
\includegraphics[width=0.5\linewidth]{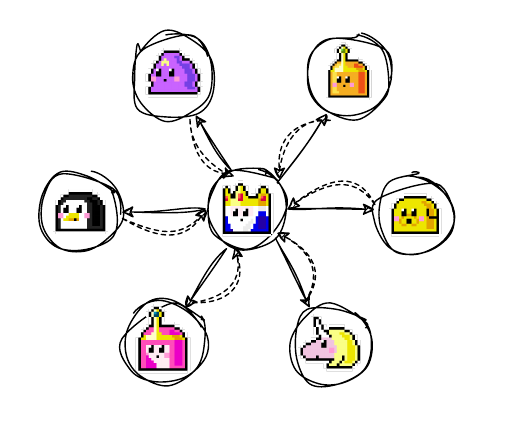}
\vspace{-0.5em}
\caption{Demonstration of \textbf{star} structure as spatial communication topology.}
\label{fig:topo_star}
\end{figure*}

The star structure resembles the tree structure and can essentially be viewed as a tree with a depth of two. When utilizing the star configuration for spatial communication, the central administrative agent receives queries from the user and dispatches instructions to subordinate agents. Upon completing their tasks using various tools, these subordinate agents return all outputs to the administrative agent, which then compiles a final summary, as illustrated in \Cref{fig:topo_star}.

\subsection{Complete Graph Structure}\label{app:topo_all} 
\begin{figure*}[!h]
\centering
\includegraphics[width=0.4\linewidth]{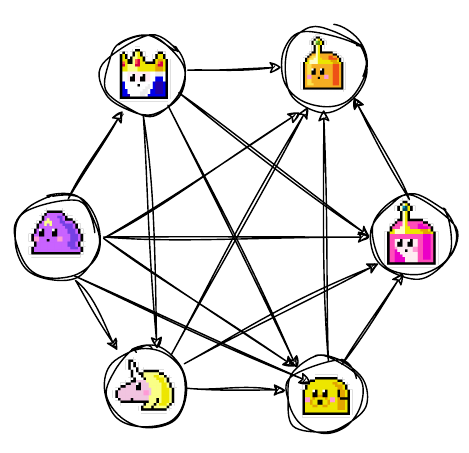}
\vspace{-0.5em}
\caption{Demonstration of \textbf{complete graph} structure as spatial communication topology.}
\label{fig:topo_complete}
\end{figure*}

In the main text, we refer to the structure shown in \Cref{fig:topo_complete} as a complete graph. However, this complete graph differs from the traditional definition, \textit{i.e.}, an undirected graph where each vertex is connected to every other vertex. Instead, it is a directed graph that would represent a complete graph if converted to an undirected form. This distinction is necessary because the execution of the multi-agent system relies on topological ordering~\citep{qian2024scaling,zhuge2024gptswarm}, requiring the spatial communication topology to be a DAG. In MacNet~\citep{qian2024scaling}, this structure is also referred to as a “Mesh graph.” After executing in the order determined by topological sorting, the final agent summarizes the dialogue and provides a concluding output or reflection.

\subsection{Layered Graph Structure}\label{app:topo_layered}
\begin{figure*}[!h]
\centering
\includegraphics[width=0.5\linewidth]{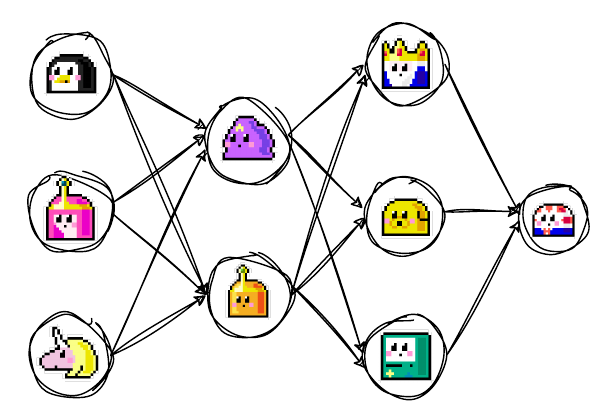}
\vspace{-0.5em}
\caption{Demonstration of \textbf{layered graph} structure as spatial communication topology.}
\label{fig:topo_layered}
\end{figure*}

A layered graph, proposed by \cite{qian2024scaling}, refers to the structure illustrated in \Cref{fig:topo_layered}, resembling a stacked configuration similar to a multilayer perceptron (MLP). The query is first provided to all agents in the first layer, whose outputs serve as prompts that, along with the query, are then fed to the agents in the second layer. The final layer consists of a single agent that receives information from the previous layer and generates the ultimate solution.

\subsection{Random Graph Structure}
\begin{figure*}[!h]
\centering
\includegraphics[width=0.4\linewidth]{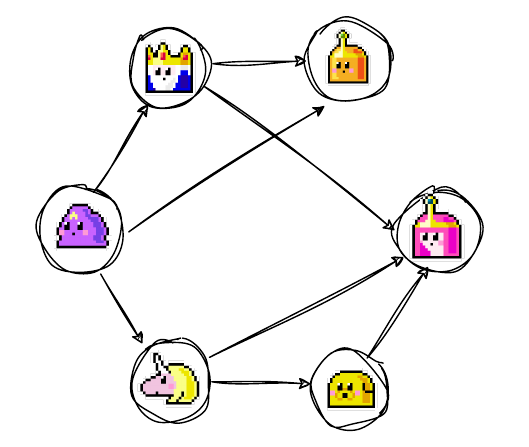}
\vspace{-0.5em}
\caption{Demonstration of \textbf{random graph} structure as spatial communication topology.}
\label{fig:topo_random}
\end{figure*}

A random graph refers to a sparse graph randomly sampled from a complete graph, as illustrated in \Cref{fig:topo_random}. Irregular random structures have been shown to outperform regular fully connected structures~\citep{qian2024scaling}, which is attributed to the presence of random edge connections. Analogous to social networks, these connections can link ``unacquainted" agents through direct shortcuts, transforming them into adjacent ``acquaintances" and implicitly reducing the average path length, thereby exhibiting small-world characteristics.

\subsection{LLM-Blender Structure}
\begin{figure*}[!h]
\centering
\includegraphics[width=0.25\linewidth]{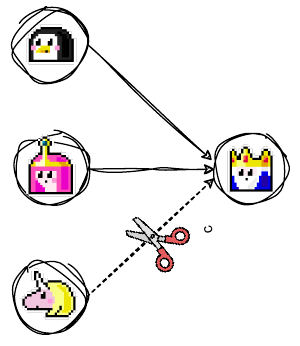}
\vspace{-0.5em}
\caption{Demonstration of \textbf{LLM-Blender} structure as spatial communication topology.}
\label{fig:topo_blender}
\end{figure*}

The structure of LLM-Blender is relatively straightforward. It feeds a query to multiple LLM-powered agents from different sources, employing a PairRanker mechanism to score each agent's output. The top $K$ responses are then selected and merged using an LLM called GenFuser, as illustrated in \Cref{fig:topo_blender}.

\section{Experimental Details}\label{app:exp}
\subsection{Baselines}\label{app:exp_baseline}
In this section, we will provide a detailed overview of the various baselines mentioned in \Cref{sec:exp_setup} and their adaptations for our evaluation.

\subsubsection{Spatial Communication Baselines}
The methods described below fall under the category of spatial communication, meaning they are designed to regulate how different agents interact and exchange information within the same dialogue round. Unless stated otherwise, we do not employ explicit inter-dialogue message passing and limit iterations to two rounds (\textit{i.e.}, $K=2$).

\paragraph{Chain} Given the diversity of benchmarks we employed (including mathematical reasoning, code generation, etc.), we organized distinct agent pools for different categories of reasoning tasks, as detailed in \Cref{app:agent_profiling}. In addition to personalized prompts for each agent, we also designed a universal prompt template applicable to all agents for generating outputs. Taking the MMLU benchmark as an example:
\begin{tcolorbox}[notitle, sharp corners, colframe=TealBlue, colback=white, 
       boxrule=3pt, boxsep=0.5pt, enhanced, 
       shadow={3pt}{-3pt}{0pt}{opacity=1,mygrey},
       title={Prompt Template for Agents on the Chain},]\label{box:prompt_chain}
\texttt{I will ask you a question and 4 answers enumerated as A, B, C and D.\\
Only one answer out of the offered 4 is correct.\\
You must choose the correct answer to the question from your perspective.\\
Using the reasoning from other agents as additional advice with critical thinking, can you give an updated answer?\\
You are strictly prohibited from imitating the analysis process of other agents.\\
Your reply must be less than 100 words but include your answer and a brief step-by-step analysis of the question.\\
The first line of your reply must contain only one letter(for example : A, B, C or D)}
\end{tcolorbox}

We utilize the output from the final agent as the decision for the entire system.

\paragraph{Star}
When employing the star structure, as described in \Cref{app:struc_star}, we designate one agent as the administrative agent and utilize the other agents as subordinates. The administrative agent ultimately collects outputs from the subordinate agents to make a decision. During the final decision-making process, the prompt is as follows:
\begin{tcolorbox}[notitle, sharp corners, colframe=TealBlue, colback=white, 
       boxrule=3pt, boxsep=0.5pt, enhanced, 
       shadow={3pt}{-3pt}{0pt}{opacity=1,mygrey},
       title={Prompt Template for Decision Making},]\label{box:prompt_decision}
\texttt{You are the top decision-maker and are good at analyzing and summarizing other people's opinions, finding errors and giving final answers.}
\end{tcolorbox}

\paragraph{Tree} We reduce the tree structure to a \textbf{binary tree} and sequentially assign agents based on the binary tree indexing, depending on the number of agents. The outputs from all non-root nodes are ultimately relayed to the root node's agent for the final decision, with the prompt template remaining the same as the Star structure.
\vspace{-0.6em}
\paragraph{Complete Graph}
We employ the structure outlined in \Cref{fig:topo_complete} and execute the input/output for each agent node through topological sorting. Before performing topological sorting, it may be necessary to apply the DAG sampling method discussed in \Cref{sec:stgraph} to ensure that the spatial communication graph is a DAG. Given the relative complexity of the graph structure, which does not possess a straightforward core agent like a chain or tree, we introduce an additional summarizer node to which all other nodes direct their outputs. Naturally, this summarizer node is executed last in the topological order, positioning it as the final decision-making expert.
\vspace{-0.6em}
\paragraph{Layered Graph} As discussed in \Cref{app:topo_layered}, we arrange the agents in an MLP-like layered structure, ensuring that the final layer contains only one agent. This agent receives outputs from all agents in the preceding layer and produces the final solution.
\vspace{-0.6em}
\paragraph{Random Graph} The implementation of a random graph is similar to that of a complete graph. It also begins with DAG sampling, followed by execution through topological sorting, and concludes with a summarizer agent that generates the overall response.
\vspace{-0.6em}
\paragraph{LLM-Blender} LLM-Blender~\citep{blender} was originally designed to consolidate responses from various LLM architectures. In this context, we treat it as a spatial message-passing paradigm, standardizing all agents to utilize either \llmname{gpt-3.5} or \llmname{gpt-4}, with the final output from GenFuser serving as the solution. Notably, LLM-Blender is specifically tailored for single-turn dialogues; thus, we do not employ multi-turn dialogues in conjunction with LLM-Blender.
\vspace{-0.6em}
\paragraph{GPTSwarm} GPTSwarm~\citep{zhuge2024gptswarm} conceptualizes the connections among all agents as a parameterized, dense adjacency matrix, which is continuously optimized to enhance collaborative performance. In the original paper, distinct internal structures were customized for different agents, such as configuring a specific agent to first receive a query, followed by performing \texttt{FileAnalyze} and \texttt{WebSearch}, and ultimately outputting the results. To ensure a fair comparison, we did not utilize such configurations in \Cref{tab:main_results}; instead, we assigned each agent different profiles, similar to the other structures mentioned above, along with possible external tools like a Python compiler or Wikipedia searcher. Our implementation is based on the resources available at \url{https://github.com/metauto-ai/GPTSwarm}. \textbf{Important Note}: in the originally open-sourced code, the multi-agent collaboration for MMLU dataset only transmitted the options A/B/C/D during dialogues, without including the content of agents' reasoning process, which is not consistent with the description in Section 2.2 of their manuscript. To ensure a fair comparison and maintain consistency with the original description, we modified their code to transmit both the choices and the reasoning process.

\subsubsection{Temporal Communication Baselines}

\begin{figure*}[!h]
\centering
\includegraphics[width=0.4\linewidth]{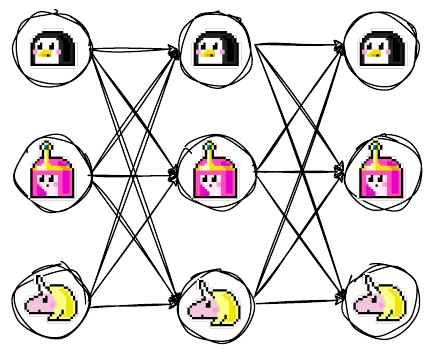}
\vspace{-0.5em}
\caption{Demonstration of \textbf{LLM-Debate} structure as temporal communication topology.}
\label{fig:topo_debate}
\end{figure*}

\paragraph{LLM-Debate} LLM-Debate~\citep{arXiv2023_MultiAgent-Debate} is designed for multiple agents to engage in a debate, where in each round, every agent receives the outputs of all agents from the previous round before making their own statements. Consequently, it essentially forms a fully connected temporal communication graph, as illustrated in \Cref{fig:topo_debate}.

\begin{figure*}[!h]
\centering
\includegraphics[width=0.5\linewidth]{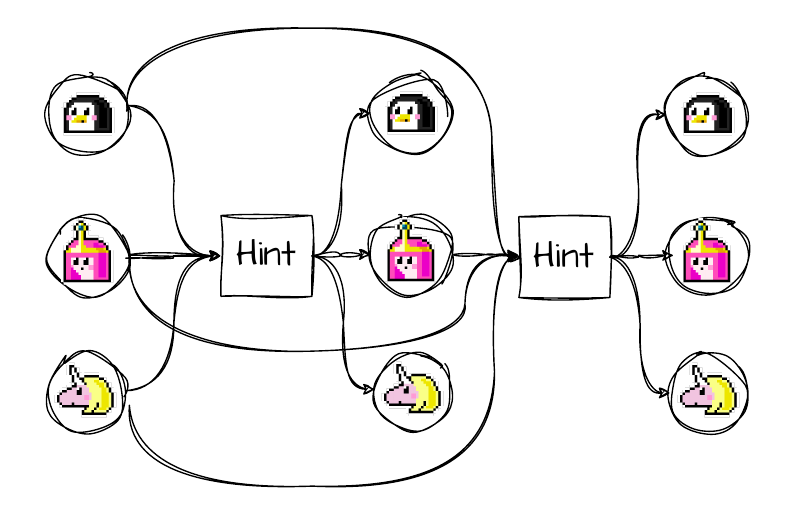}
\vspace{-0.5em}
\caption{Demonstration of \textbf{PHP} structure as temporal communication topology.}
\label{fig:topo_php}
\end{figure*}

\paragraph{PHP} PHP~\citep{PHPrompting} progressively improves prompts by utilizing the entirety of historical dialogue, offering potential "hint" prompts. In this context, we adapt this setting to the multi-agent collaboration framework, using the decisions made after each round of dialogue as hint prompts for all agents in the subsequent round, as depicted in \Cref{fig:topo_php}.

\paragraph{DyLAN} DyLAN~\citep{arXiv2023_Dynamic-LLM-Agent} primarily focuses on optimizing temporal communication and reducing redundancy by employing a specific scoring mechanism to eliminate low-quality outputs between every two rounds of dialogue. We utilize the official implementation available at \url{https://github.com/SALT-NLP/DyLAN}.

\subsubsection{Others}
\begin{figure*}[!t]
\vspace{-0.6em}
\centering
\includegraphics[width=0.8\linewidth]{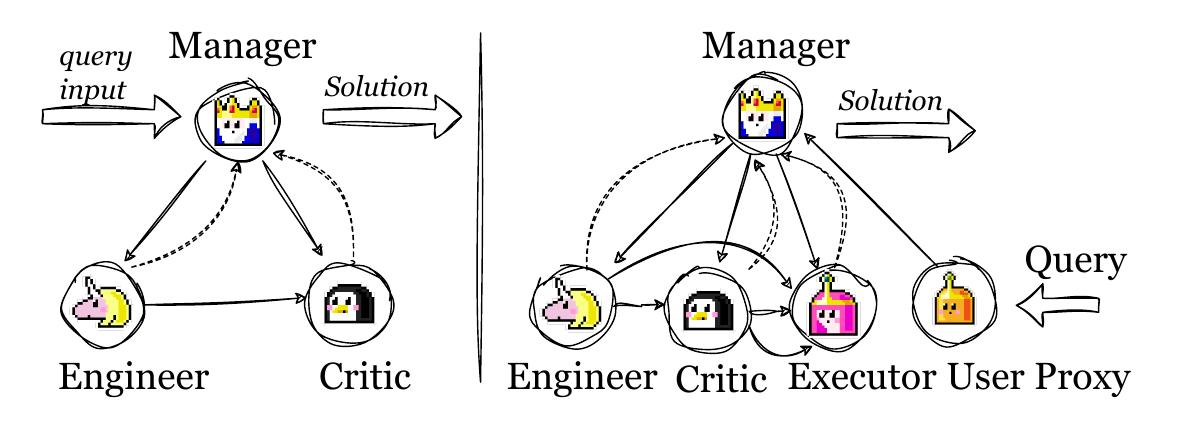}
\vspace{-1.em}
\caption{The AutoGen system design with three/five LLM-based agents. The dashed line indicates the agent will propagate its rationale or output back to the manager for its final decision-making.}
\label{fig:autogen}
\vspace{-1.em}
\end{figure*}

In \Cref{sec:plugin}, we integrated \ourmethod with AutoGen and GPTSwarm under both three-agent and five-agent configurations. Here, we elaborate on how we implemented AutoGen, which is inherently a customizable framework. Based on the setup from AutoGen (A5: Dynamic Group Chat), we define five roles: user proxy, manager, engineer, critic, and code executor. For the three-agent configuration, we condense these roles into manager, engineer, and critic. The detailed communication structure is depicted in \Cref{fig:autogen}.

\subsection{Agent Profiling}\label{app:agent_profiling}
Previous works~\citep{NAACL2024_Agent-Self-Collaboration} have formally established that assigning different personas or roles to LLM-based agents can enhance cognitive synergy among agents. Consequently, we utilized \llmname{gpt-4} to generate a series of agent profiles for various tasks, thereby promoting diversity and collective intelligence in a multi-agent setting. 

\subsubsection{Profile Examples for General Reasoning}
Below are some examples of agent profiles tailored for \textbf{general reasoning} tasks:

\begin{tcolorbox}[notitle, sharp corners, colframe=Periwinkle, colback=white, 
       boxrule=3pt, boxsep=0.5pt, enhanced, 
       shadow={3pt}{-3pt}{0pt}{opacity=1,mygrey},
       title={Knowledge Expert},]\label{box:prompt_mmlu_knowle}
       \small
\texttt{You are a knowledgeable expert in question answering.\\
Please give several key entities that need to be searched in Wikipedia to solve the problem. \\
Key entities that need to be searched are included between two '@' when output, for example: @catfish effect@, @broken window effect@, @Shakespeare@.\\
If there is no entity in the question that needs to be searched in Wikipedia, you don't have to provide it}
\end{tcolorbox}

\begin{tcolorbox}[notitle, sharp corners, colframe=Periwinkle, colback=white, 
       boxrule=3pt, boxsep=0.5pt, enhanced, 
       shadow={3pt}{-3pt}{0pt}{opacity=1,mygrey},
       title={Wiki Searcher},]\label{box:prompt_mmlu_wiki}
       \small
\texttt{You will be given a question and a Wikipedia overview of the key entities within it.\\
Please refer to them step by step to give your answer.\\
And point out potential issues in other agent's analysis.}
\end{tcolorbox}

\begin{tcolorbox}[notitle, sharp corners, colframe=Periwinkle, colback=white, 
       boxrule=3pt, boxsep=0.5pt, enhanced, 
       shadow={3pt}{-3pt}{0pt}{opacity=1,mygrey},
       title={Critic},]\label{box:prompt_mmlu_critic}
       \small
\texttt{You are an excellent critic.\\
Please point out potential issues in other agent's analysis point by point.}
\end{tcolorbox}

\begin{tcolorbox}[notitle, sharp corners, colframe=Periwinkle, colback=white, 
       boxrule=3pt, boxsep=0.5pt, enhanced, 
       shadow={3pt}{-3pt}{0pt}{opacity=1,mygrey},
       title={Mathematician},]\label{box:prompt_mmlu_math}
       \small
\texttt{You are a mathematician who is good at math games, arithmetic calculation, and long-term planning.
}
\end{tcolorbox}

\begin{tcolorbox}[notitle, sharp corners, colframe=Periwinkle, colback=white, 
       boxrule=3pt, boxsep=0.5pt, enhanced, 
       shadow={3pt}{-3pt}{0pt}{opacity=1,mygrey},
       title={Programmer},]\label{box:prompt_mmlu_prog}
       \small 
\texttt{
You are good at computer science, engineering, and physics.\\
You have experience in designing and developing computer software and hardware.\\
You are especially good at writing code or complex programs with Python, C++, MATLAB, JAVA, etc.
}
\end{tcolorbox}

\begin{tcolorbox}[notitle, sharp corners, colframe=Periwinkle, colback=white, 
       boxrule=3pt, boxsep=0.5pt, enhanced, 
       shadow={3pt}{-3pt}{0pt}{opacity=1,mygrey},
       title={Doctor},]\label{box:prompt_mmlu_doc}
       \small
\texttt{
You are a doctor and come up with creative treatments for illnesses or diseases.\\
You are able to recommend conventional medicines, herbal remedies and other natural alternatives. \\
You also consider the patient's age, lifestyle and medical history when providing your recommendations.
}
\end{tcolorbox}

\begin{tcolorbox}[notitle, sharp corners, colframe=Periwinkle, colback=white, 
       boxrule=3pt, boxsep=0.5pt, enhanced, 
       shadow={3pt}{-3pt}{0pt}{opacity=1,mygrey},
       title={Economist},]\label{box:prompt_mmlu_econ}
       \small
\texttt{
You are good at economics, finance, and business.\\
You have experience on understanding charts while interpreting the macroeconomic environment prevailing across world economies.
}
\end{tcolorbox}

\subsubsection{Profile Examples for Mathematical Reasoning}
Below are some examples of agent profiles tailored for \textbf{mathematical reasoning} tasks:

\begin{tcolorbox}[notitle, sharp corners, colframe=ForestGreen, colback=white, 
       boxrule=3pt, boxsep=0.5pt, enhanced, 
       shadow={3pt}{-3pt}{0pt}{opacity=1,mygrey},
       title={Math Solver},]\label{box:prompt_math_math}
       \small
\texttt{
You are a math expert.\\
You will be given a math problem and hints from other agents. \\
Give your own solving process step by step based on hints. \\
The last line of your output contains only the final result without any units, for example: The answer is 140.\\
You will be given some examples you may refer to.
}
\end{tcolorbox}

\begin{tcolorbox}[notitle, sharp corners, colframe=ForestGreen, colback=white, 
       boxrule=3pt, boxsep=0.5pt, enhanced, 
       shadow={3pt}{-3pt}{0pt}{opacity=1,mygrey},
       title={Mathematical Analyst},]\label{box:prompt_math_maha}
       \small
\texttt{
You are a mathematical analyst. \\
You will be given a math problem, analysis and code from other agents. \\
You need to first analyze the problem-solving process step by step, where the variables are represented by letters. \\
Then you substitute the values into the analysis process to perform calculations and get the results.\\
The last line of your output contains only the final result without any units, for example: The answer is 140\\
You will be given some examples you may refer to.
}
\end{tcolorbox}

\begin{tcolorbox}[notitle, sharp corners, colframe=ForestGreen, colback=white, 
       boxrule=3pt, boxsep=0.5pt, enhanced, 
       shadow={3pt}{-3pt}{0pt}{opacity=1,mygrey},
       title={Programming Expert},]\label{box:prompt_math_prog}
       \small
\texttt{
You are a programming expert. \\
You will be given a math problem, analysis and code from other agents. 
Integrate step-by-step reasoning and Python code to solve math problems. \\
Analyze the question and write functions to solve the problem. \\
The function should not take any arguments and use the final result as the return value. \\
The last line of code calls the function you wrote and assigns the return value to the \\(answer\\) variable. \\
Use a Python code block to write your response. For example: <some python code>\\
Do not include anything other than Python code blocks in your response.
You will be given some examples you may refer to.\\
}
\end{tcolorbox}

\subsubsection{Profile Examples for Code Generarion}
Below are some examples of agent profiles tailored for \textbf{code generation} tasks:

\begin{tcolorbox}[notitle, sharp corners, colframe=YellowGreen, colback=white, 
       boxrule=3pt, boxsep=0.5pt, enhanced, 
       shadow={3pt}{-3pt}{0pt}{opacity=1,mygrey},
       title={Project Manager},]\label{box:prompt_code_porj}
       \small
\texttt{
"You are a project manager. "\\
"You will be given a function signature and its docstring by the user. "\\
"You are responsible for overseeing the overall structure of the code, ensuring that the code is structured to complete the task Implement code concisely and correctly without pursuing over-engineering."\\
"You need to suggest optimal design patterns to ensure that the code follows best practices for maintainability and flexibility. "\\
"You can specify the overall design of the code, including the classes that need to be defined(maybe none) and the functions used (maybe only one function) ."\\
"I hope your reply will be more concise. Preferably within fifty words. Don’t list too many points."
}
\end{tcolorbox}

\begin{tcolorbox}[notitle, sharp corners, colframe=YellowGreen, colback=white, 
       boxrule=3pt, boxsep=0.5pt, enhanced, 
       shadow={3pt}{-3pt}{0pt}{opacity=1,mygrey},
       title={Algorithm Designer},]\label{box:prompt_code_algo}
       \small
\texttt{
"You are an algorithm designer. "\\
"You will be given a function signature and its docstring by the user. "\\
"You need to specify the specific design of the algorithm, including the classes that may be defined and the functions used. "\\
"You need to generate the detailed documentation, including explanations of the algorithm, usage instructions, and API references. "\\
"When the implementation logic is complex, you can give the pseudocode logic of the main algorithm."\\
"I hope your reply will be more concise. Preferably within fifty words. Don’t list too many points.",
}
\end{tcolorbox}

\begin{tcolorbox}[notitle, sharp corners, colframe=YellowGreen, colback=white, 
       boxrule=3pt, boxsep=0.5pt, enhanced, 
       shadow={3pt}{-3pt}{0pt}{opacity=1,mygrey},
       title={Programming Expert},]\label{box:prompt_code_expert}
       \small
\texttt{
"You are a programming expert. "\\
"You will be given a function signature and its docstring by the user. "\\
"You may be able to get the output results of other agents. They may have passed internal tests, but they may not be completely correct. "
"Write your full implementation (restate the function signature). "\\
"Use a Python code block to write your response. For example: <some python code>\\
"Do not include anything other than Python code blocks in your response. "\\
"Do not change function names and input variable types in tasks."
}
\end{tcolorbox}

\begin{tcolorbox}[notitle, sharp corners, colframe=YellowGreen, colback=white, 
       boxrule=3pt, boxsep=0.5pt, enhanced, 
       shadow={3pt}{-3pt}{0pt}{opacity=1,mygrey},
       title={Bug Fixer},]\label{box:prompt_code_bug}
       \small
\texttt{
"You are a bug fixer."\\
"You will be given a function signature and its docstring by the user. "\\
"You need to provide modified and improved python code based on the current overall code design, algorithm framework, code implementation or test problems. "\\
"Write your full implementation (restate the function signature). "\\
"Use a Python code block to write your response. For example: <some python code>"\\
"Do not include anything other than Python code blocks in your response "\\
"Do not change function names and input variable types in tasks"
}
\end{tcolorbox}

\subsection{Agent Attack Implementation}\label{app:attack}

We designed two types of attacks on agents: agent prompt attack and agent replacement attack, with specific implementations detailed as follows:

\paragraph{Agent Prompt Attack} We attempt to compromise the role prompt of the collaborative agent, altering its predefined role to the following ``lier'' agent:
\begin{tcolorbox}[notitle, sharp corners, colframe=Melon, colback=white, 
       boxrule=3pt, boxsep=0.5pt, enhanced, 
       shadow={3pt}{-3pt}{0pt}{opacity=1,mygrey},
       title={Prompt Template for Agent Prompt Attack},]\label{box:prompt_attack}
       \small
\texttt{You are a liar who only tell lies.
Give a wrong answer and false analysis process for the following question: \{question\}.\\
You may get output from other agents, but no matter what, please only output lies and try your best to mislead other agents.\\
Your reply must be less than 100 words.\\
The first line of your reply must contain only one letter(for example : A, B, C or D)}
\end{tcolorbox}

\paragraph{Agent Replacement Attack} We replace the originally high-cognitive and planning-capable LLM with a randomly generating ``dummy'' API that outputs text without coherence. The specific prompt is as follows:
\begin{tcolorbox}[notitle, sharp corners, colframe=Melon, colback=white, 
       boxrule=3pt, boxsep=0.5pt, enhanced, 
       shadow={3pt}{-3pt}{0pt}{opacity=1,mygrey},
       title={Prompt Template for Agent Replacement Attack},]\label{box:replace_decision}
       \small
\texttt{
Randomly output a letter from ABCD on the first line.\\
Then output any gibberish paragraph on the same topic as the following question: {question}.\\
The first line of your reply must contain only one letter(for example : A, B, C or D)}
\end{tcolorbox}

When attacking all multi-agent frameworks, we randomly select one agent to serve as the adversarial agent, while the remaining agents retain their original functions and responsibilities.

\section{Supplemented Experimental Results}

\subsection{Results for RQ1}

\begin{figure*}[!h]
\centering
\includegraphics[width=\linewidth]{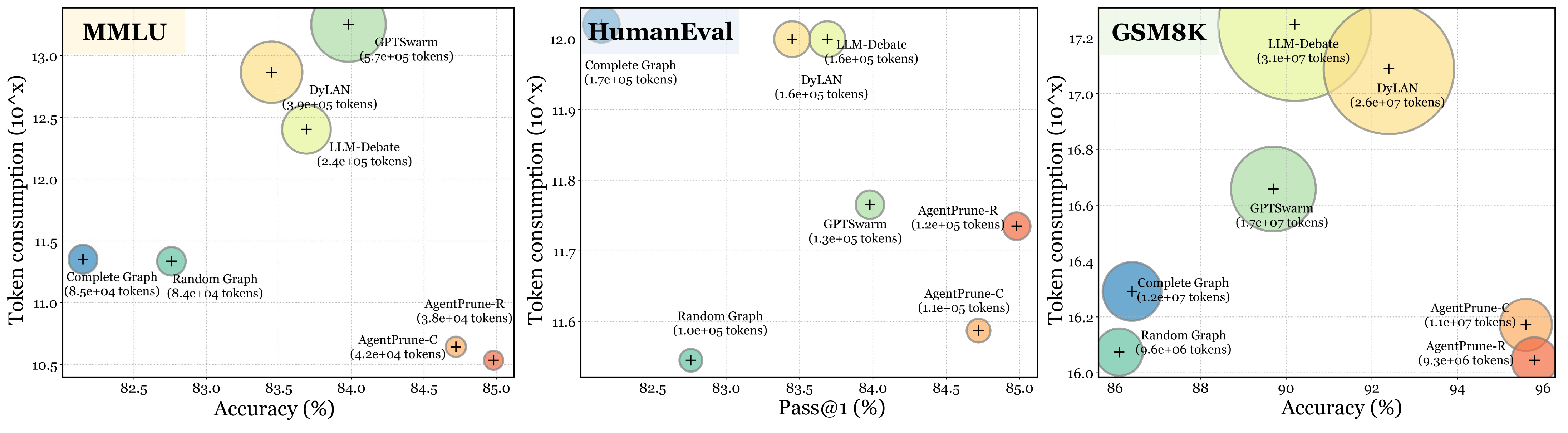}
\caption{Scatter plot illustrating the performance metrics and \textbf{total token consumption} of different multi-agent communication topologies across the MMLU, HumanEval, and GSM8K datasets. The diameter of each point is proportional to the value on the y-axis, representing token consumption.}
\label{fig:scatter_2}
\end{figure*}

\begin{figure*}[!h]
\centering
\includegraphics[width=\linewidth]{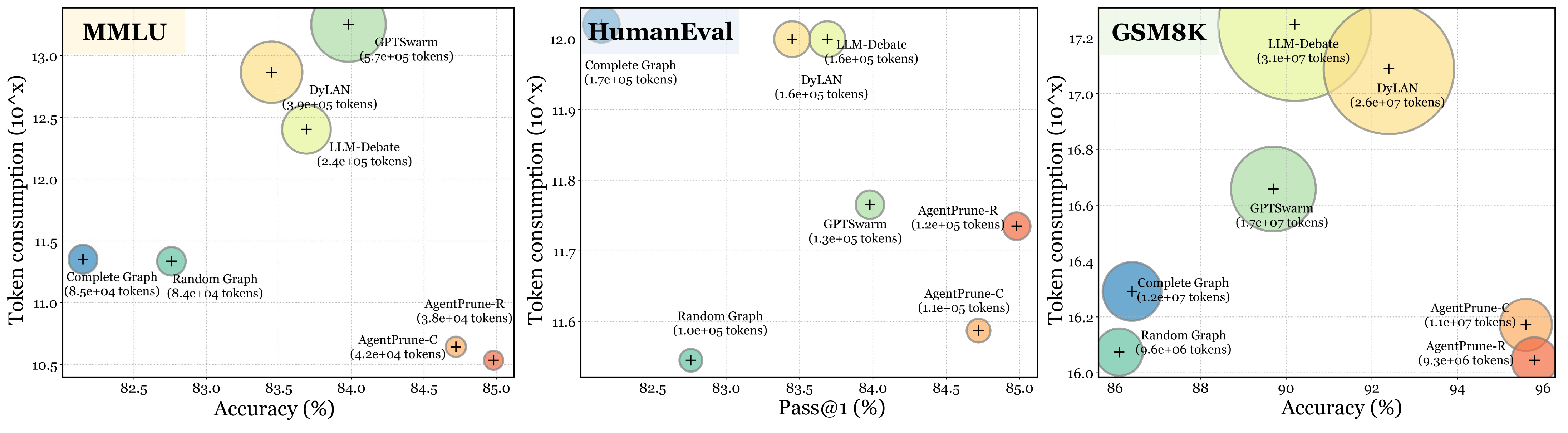}
\caption{Scatter plot illustrating the performance metrics and \textbf{completion token consumption} of different multi-agent communication topologies across the MMLU, HumanEval, and GSM8K datasets. The diameter of each point is proportional to the value on the y-axis, representing token consumption.}
\label{fig:scatter_3}
\end{figure*}

\begin{figure*}[!h]
\centering
\includegraphics[width=\linewidth]{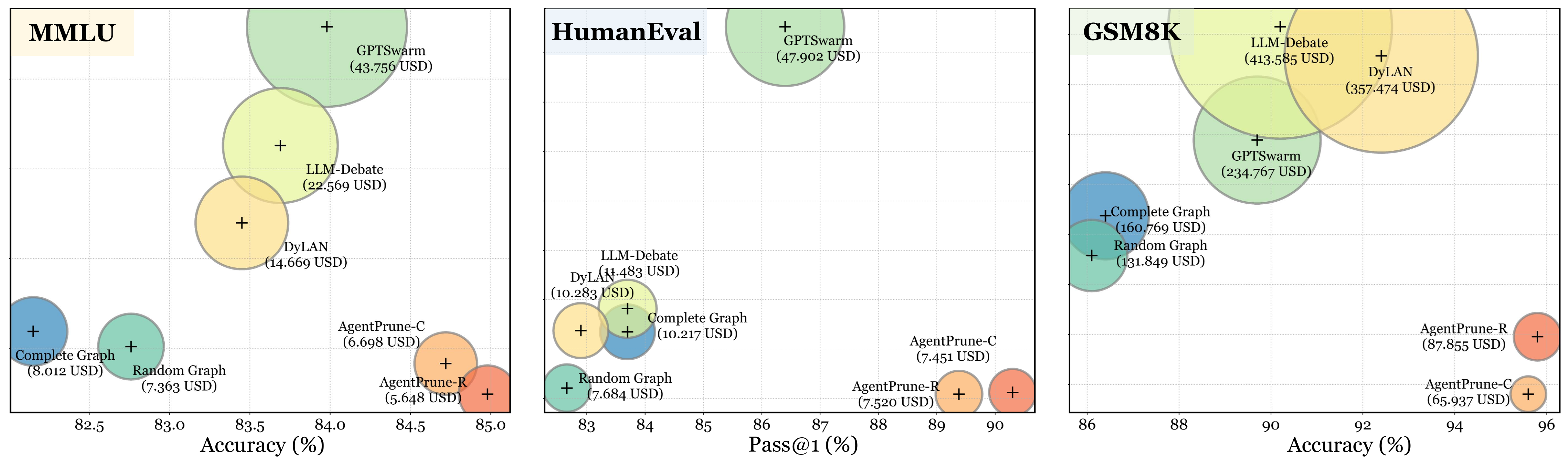}
\caption{Scatter plot illustrating the performance metrics and \textbf{total cost (USD)} of different multi-agent communication topologies across the MMLU, HumanEval, and GSM8K datasets. The diameter of each point is proportional to the value on the y-axis, representing the economic cost.}
\label{fig:scatter_4}
\end{figure*}

\Cref{fig:scatter_2,fig:scatter_3,fig:scatter_4} compares \ourmethod with other communication topologies in terms of completion tokens, overall tokens, and overall cost (USD). Notably, across multiple datasets, \ourmethod achieves superior performance at a fraction of the cost, often as low as one-half or even one-tenth of the economic expense of SOTA topologies. For instance, on the MMLU dataset, \ourmethod surpasses GPTSwarm’s performance with a cost of only $\$5.6$, compared to GPTSwarm's $\$43.56$. Similarly, on the GSM8K dataset, \ourmethod outperforms DyLAN with a cost of $\$65.9$, whereas DyLAN incurs a cost of $\$357.47$.

\subsection{Results for RQ2}

\Cref{tab:combine_cost_3} presents a comparison of the performance, prompt/completion token consumption, and cost between \ourmethod integrated with three-agent-based AutoGen and GPTSwarm frameworks.

\begin{table}[t!]
  \centering
  \caption{
  \textbf{Performance and Cost Comparison before and after combining \ourmethod.} We evaluated the performance and economic cost comparison of \ourmethod in conjunction with two classic multi-agent systems, AutoGen and GPTSwarm, under a \textbf{three} \llmname{gpt-4}-based setting.
  }
  \label{tab:combine_cost_3}
  \vspace{-0.5em}
  \renewcommand\tabcolsep{6.5pt}
  \renewcommand\arraystretch{1.1}
  \footnotesize 
  \begin{tabular}{lc|cccc} 
    \Xhline{1.2pt}
    \rowcolor{CadetBlue!20} 
    \textbf{Dataset} & \textbf{Method} & \textbf{Performance} & \textbf{\# Prompt Tokens}  & \textbf{\# Completion Tokens} & \textbf{Cost}\\
    \Xhline{1pt}
    \multirow{2}{*}{MMLU} & AutoGen & $81.94$ & $346,028$ & $66,204$ & $\$5.446$ \\
    & +\ourmethod & $82.20({\color{RedOrange} \uparrow 0.26})$ & $274,665(\color{RedOrange} 79.4\%)$  & $66,803$ & $\$4.750$\\
    \hline
     \rowcolor{gray!10}& AutoGen & $83.66$ & $351,985$ & $90,762$ & $\$6.242$ \\
   \rowcolor{gray!10} \multirow{-2}{*}{HumanEval}& +\ourmethod & $85.04({\color{RedOrange} \uparrow 1.38})$ & $254,203({\color{orange} 72.2\%})$ & $89,362$ & $\$5.282$\\
    \hline
     & AutoGen & $87.23$ & $2,317,937$ & $624,055$ & $\$41.92$  \\
    \multirow{-2}{*}{GSM8K}& +\ourmethod & $88.51({\color{RedOrange} \uparrow 1.23})$ & $1,481,780 ({\color{orange} 63.9\%})$ & $629,771$ & $\$33.71$\\
    \hline
    \rowcolor{gray!10} & GPTSwarm & $83.32$ & $1,521,504$ & $325,994$ & $\$24.99$ \\
    \rowcolor{gray!10}\multirow{-2}{*}{MMLU}& +\ourmethod & $83.66({\color{RedOrange} \uparrow 0.34})$ & $554,698({\color{orange} 35.8\%})$ & $336,887$ & $\$15.65$\\
    \hline
    
     & GPTSwarm & $83.62$ & $1,478,312$ & $612,815$ & $\$33.16$\\
    \multirow{-2}{*}{HumanEval}& +\ourmethod & $84.74({\color{RedOrange} \uparrow 1.12})$ & $432,480({\color{orange} 29.2\%})$ & $598,367$ & $\$22.07$\\
    \hline
    \rowcolor{gray!10} & GPTSwarm & $87.85$ & $6,274,665$ & $186,510$ & $\$68.34$ \\
    \rowcolor{gray!10}\multirow{-2}{*}{GSM8K}& +\ourmethod & $88.30({\color{RedOrange} \uparrow 0.45})$ & $3,009,115({\color{orange} 47.9\%})$ & $173,296$ & $\$35.29$\\
    \hline
    \Xhline{1.2pt}
  \end{tabular}
\end{table}

\subsection{Results for RQ3}

\begin{figure*}[!t]
\vspace{-0.6em}
\centering
\includegraphics[width=\linewidth]{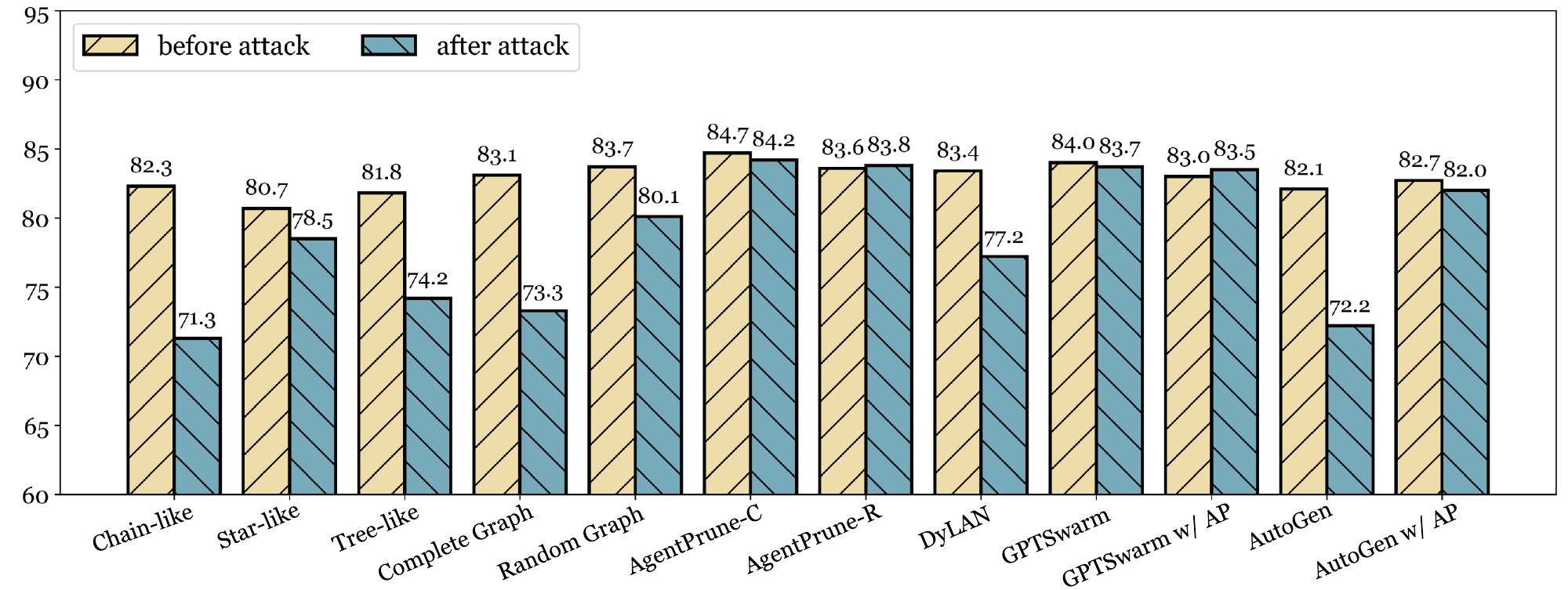}
\vspace{-1.7em}
\caption{\textbf{Performance under adversarial attack.} We compare the performance of various multi-agent frameworks before and after \textit{agent replacement attacks}. ``w/ AP'' indicates the integration with \ourmethod.}
\label{fig:attack_2}
\end{figure*}

We supplement the performance of various topologies before and after being perturbed by agent replacement attack in \Cref{fig:attack_2}.

\subsection{Results for RQ4}

\subsubsection{Ablation Study}\label{app:ablation}
We develop two variants of \ourmethod: ``\ourmethod w/o profile'', which assigns no unique roles to each agent, and ``\ourmethod w/o low-rank'', which does not implement the low-rank regularization described in \Cref{eq:rank_1}. The results are presented in \Cref{tab:ablation}.

\begin{table}[t!]
  \centering
  \caption{
  \textbf{Abltion study of \ourmethod.}
  ``\textit{w/o profile}'' denotes not assigning agents with different roles and profiles; ``\textit{w/o low-rank}'' denotes not using the low-rank regularization as described in \Cref{eq:rank_1}.
  }
  \label{tab:ablation}
  \vspace{-0.5em}
  \renewcommand\tabcolsep{6.5pt}
  \renewcommand\arraystretch{1.1}
  \footnotesize 
  \begin{tabular}{l|cccccc} 
    \Xhline{1.2pt}
\rowcolor{CadetBlue!20} 
{\textbf{Variant}}  & \textbf{MMLU} & \textbf{GSM8K} & \textbf{MultiArith} & \textbf{SVAMP} & \textbf{AQuA} & \textbf{HumanEval} \\
    \Xhline{1pt}
   \ourmethod-C & {84.72} & {95.62} & {97.25} & {91.85} & {79.47}& {89.38} \\
    \rowcolor{gray!10}\textit{w/o profile} & 84.3 &  93.7 & 96.2 & 91.7 & 79.1 & 87.8\\
    \textit{w/o low-rank} & 84.6 &  94.5 & 96.8 & 91.1 & 79.5 & 88.9\\
    \hline
   \rowcolor{gray!10}\ourmethod-R & 83.94 & {95.83} & 96.30 & {91.68}& {78.60} & {90.30}  \\
   \textit{w/o profile}  & 83.3 & {95.6} & 95.7 & {91.7}& {78.7} & {88.6} \\
   \rowcolor{gray!10}\textit{w/o low-rank}& 83.5 & {95.4} & 96.0 & {91.3}& {78.4} & {89.3}  \\
    \Xhline{1.2pt}
  \end{tabular}
  \vspace{-0.3em}
\end{table}

\subsubsection{Sensitivity Analysis}\label{app:sensitivity}
\begin{figure*}[!t]
\vspace{-0.6em}
\centering
\includegraphics[width=\linewidth]{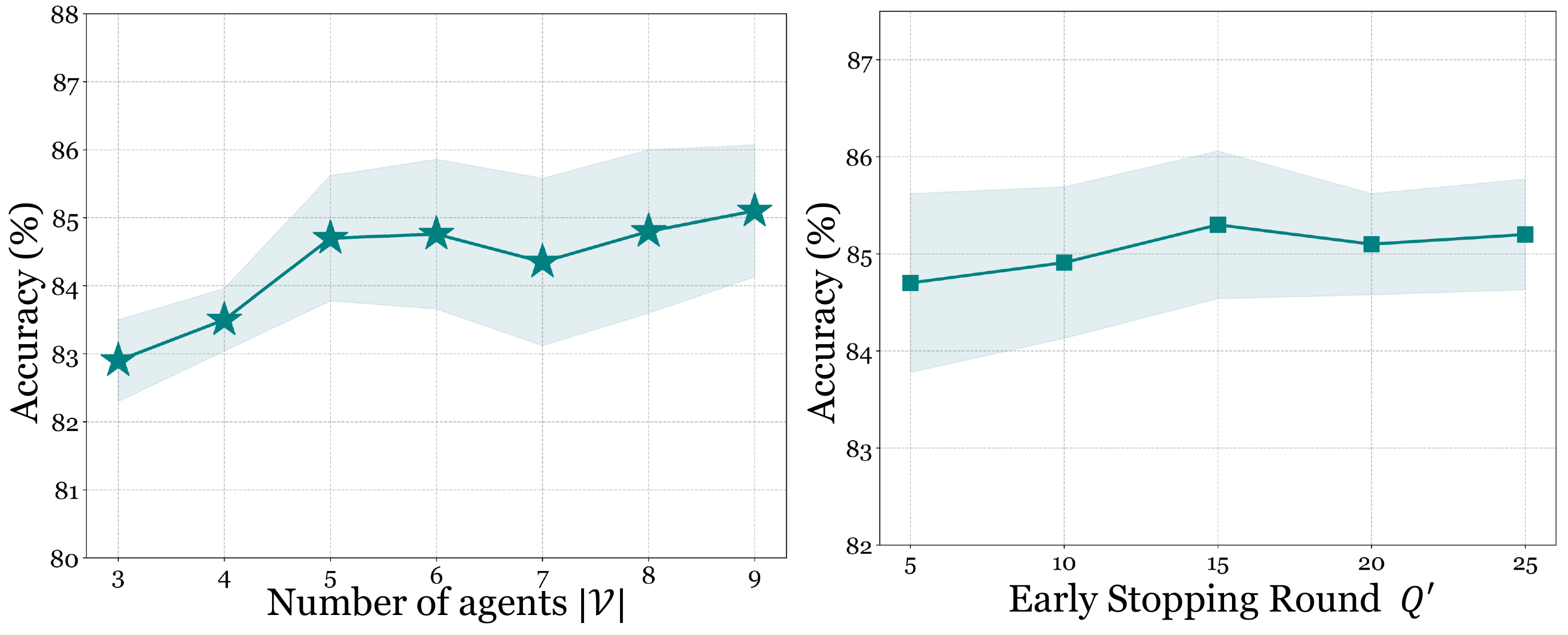}
\vspace{-1.7em}
\caption{\textbf{Parameter sensitivity analysis on \ourmethod.} we vary the number of agents $|\mathcal{V}|\in\{3,4,5,6,7,8,9\}$ and the early stopping round $Q'\in\{5,10,15,15,20,25\}$ on MMLU dataset.}
\label{fig:sensi}
\end{figure*}

We analyze the sensitivity of \ourmethod to two parameters: the number of agents \( |\mathcal{V}| \) and the early stopping round \( K \). The findings are presented in \Cref{fig:sensi}. Notably, as the number of agents increases, there is a significant performance enhancement initially (from 3 to 5 agents), while subsequent improvements (from 5 to 9 agents) in accuracy become relatively marginal. The early stopping round \( Q' \) indicates the number of queries used to optimize spatial-temporal connectivity in a multi-query training setting. Intuitively, increasing the number of optimization rounds leads to a more refined and accurate graph mask, resulting in substantial performance gains from multi-agent collaboration, along with reduced performance fluctuations. To balance performance with token savings, we consistently set \( Q' \in \{5, 10\} \).

\section{Case Study}\label{app:case_study}

\subsection{Spatial Topology Visualization}

In this section, we will demonstrate how \ourmethod operates on the predefined spatial communication structure and the resulting structure after one-shot pruning. It is important to note that the pruned sparse graph we present is likely not a Directed Acyclic Graph (DAG). This is because, during the practical application, we still need to utilize the \texttt{DAGSampling} function on the pruned graph to ensure it conforms to DAG properties before sequentially executing agent I/O.

\begin{figure*}[!h]
\centering
\includegraphics[width=0.9\linewidth]{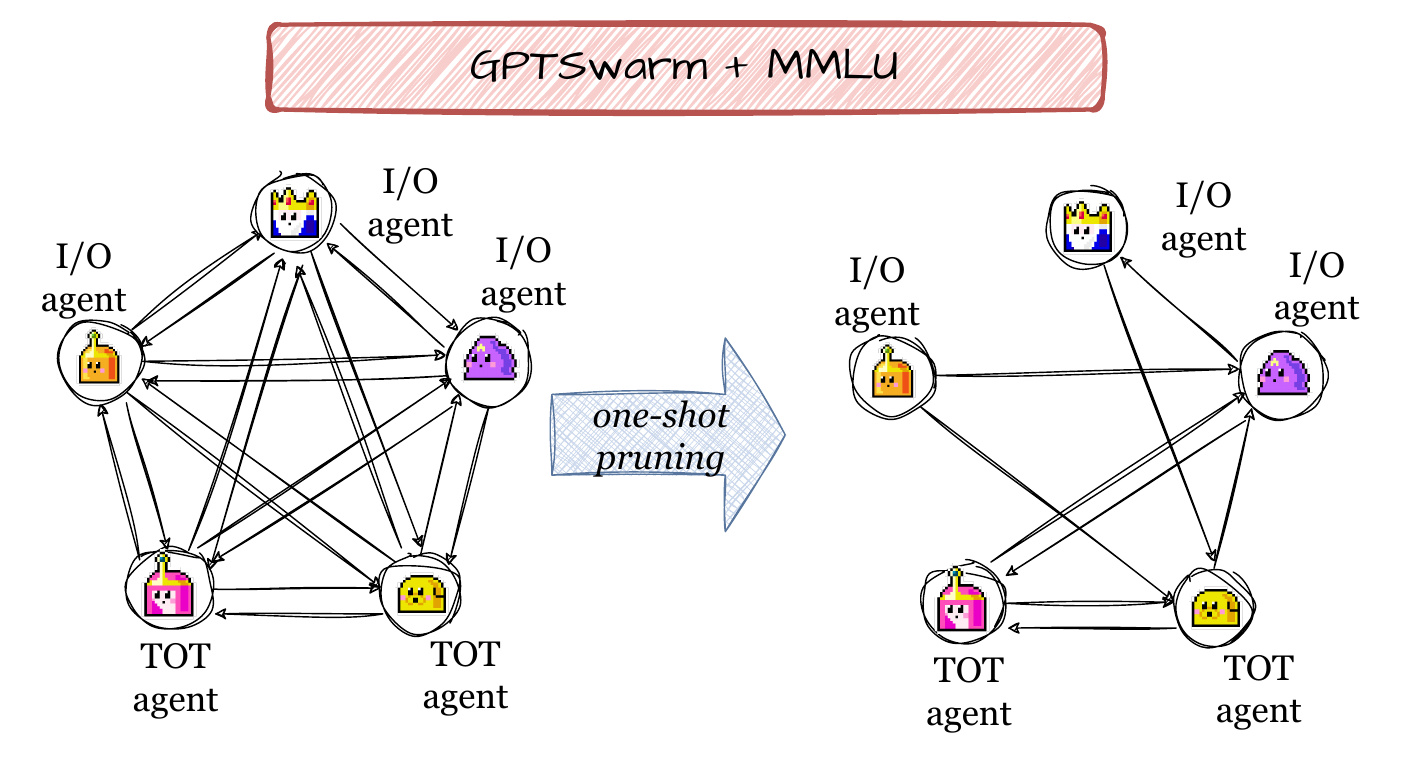}
\caption{The pruning process of \ourmethod on a five-agent GPTSwarm framework when tested on MMLU.}
\label{fig:case_swarm_mmlu}
\end{figure*}

\paragraph{GPTSwarm + MMLU} \Cref{fig:case_swarm_mmlu} illustrates how \ourmethod applies one-shot pruning within a five-agent GPTSwarm framework, where three agents serve as simple I/O agents, and the remaining two are TOT (Tree of Thoughts) agents. Notably, \ourmethod prunes many in-edges for the I/O agents, while preserving many for the TOT agents. This behavior likely stems from the stronger reasoning capabilities of TOT agents, which makes them better suited for synthesizing the agents' discussions and providing the final solution.

\begin{figure*}[!h]
\centering
\includegraphics[width=0.9\linewidth]{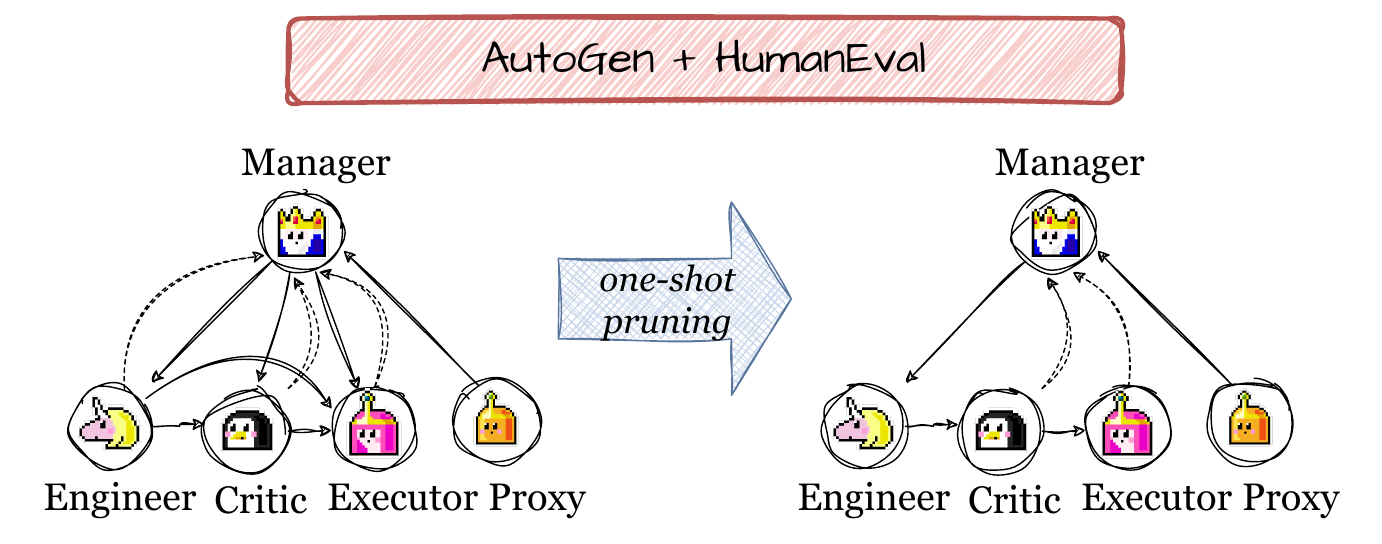}
\caption{The pruning process of \ourmethod on a five-agent AutoGen framework when tested on MMLU.}
\label{fig:case_autogen_mmlu}
\end{figure*}

\paragraph{AutoGen + HumanEval} \Cref{fig:case_autogen_mmlu} illustrates how \ourmethod applies one-shot pruning within a five-agent AutoGen framework.

\begin{figure*}[!h]
\centering
\includegraphics[width=0.9\linewidth]{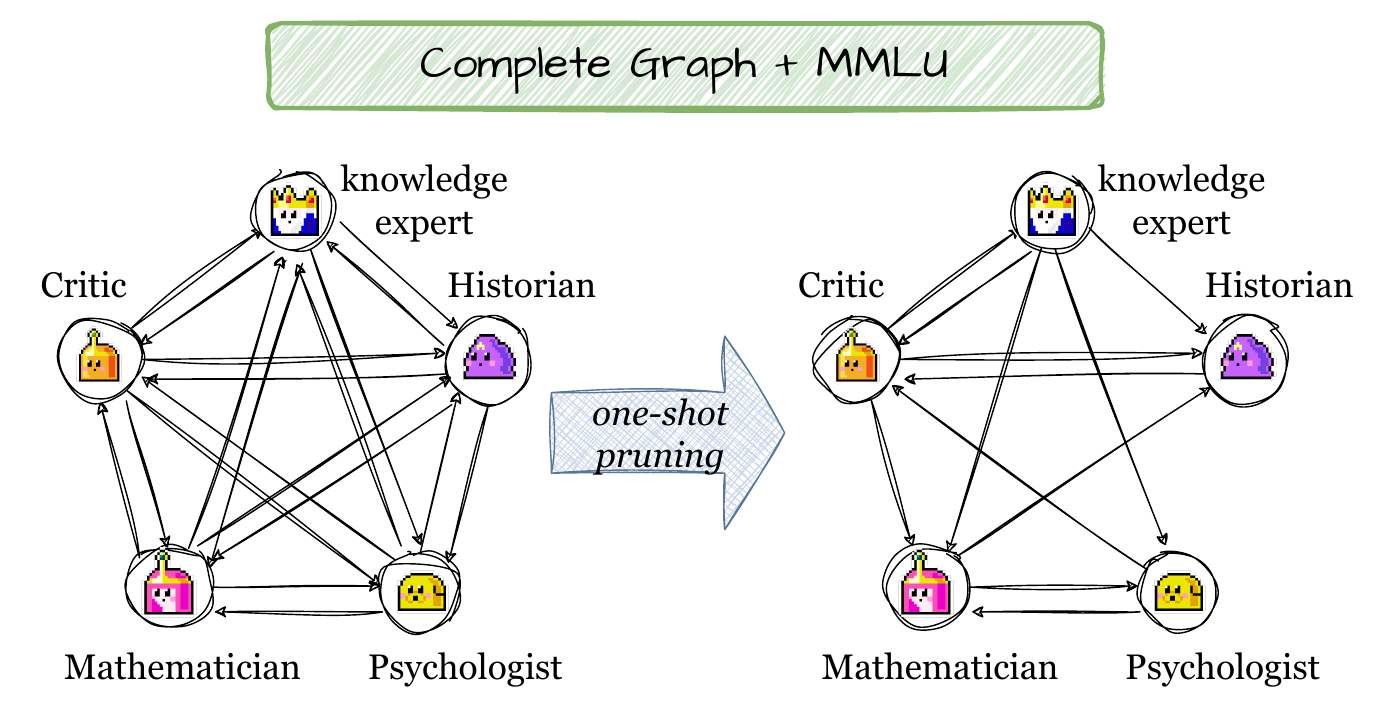}
\caption{The pruning process of \ourmethod on a five-agent complete-graph-like framework when tested on MMLU dataset.}
\label{fig:case_complete_mmlu}
\end{figure*}

\paragraph{Complete Graph + MMLU} \Cref{fig:case_complete_mmlu} illustrates how \ourmethod performs one-shot pruning to achieve a compact sparse spatial communication graph, given five predefined agent roles: knowledge expert, critic, historian, mathematician, and psychologist, along with a predefined \textbf{complete graph} structure on \textbf{MMLU} dataset. It is evident that in the pruned graph, the Critic has a high number of incoming edges, while the Knowledge Expert has a high number of outgoing edges. This aligns with their respective functions: the Critic is expected to receive a broad range of external information and provide feedback, whereas the Knowledge Expert should output useful knowledge based on their knowledge base.

\begin{figure*}[!h]
\centering
\includegraphics[width=0.9\linewidth]{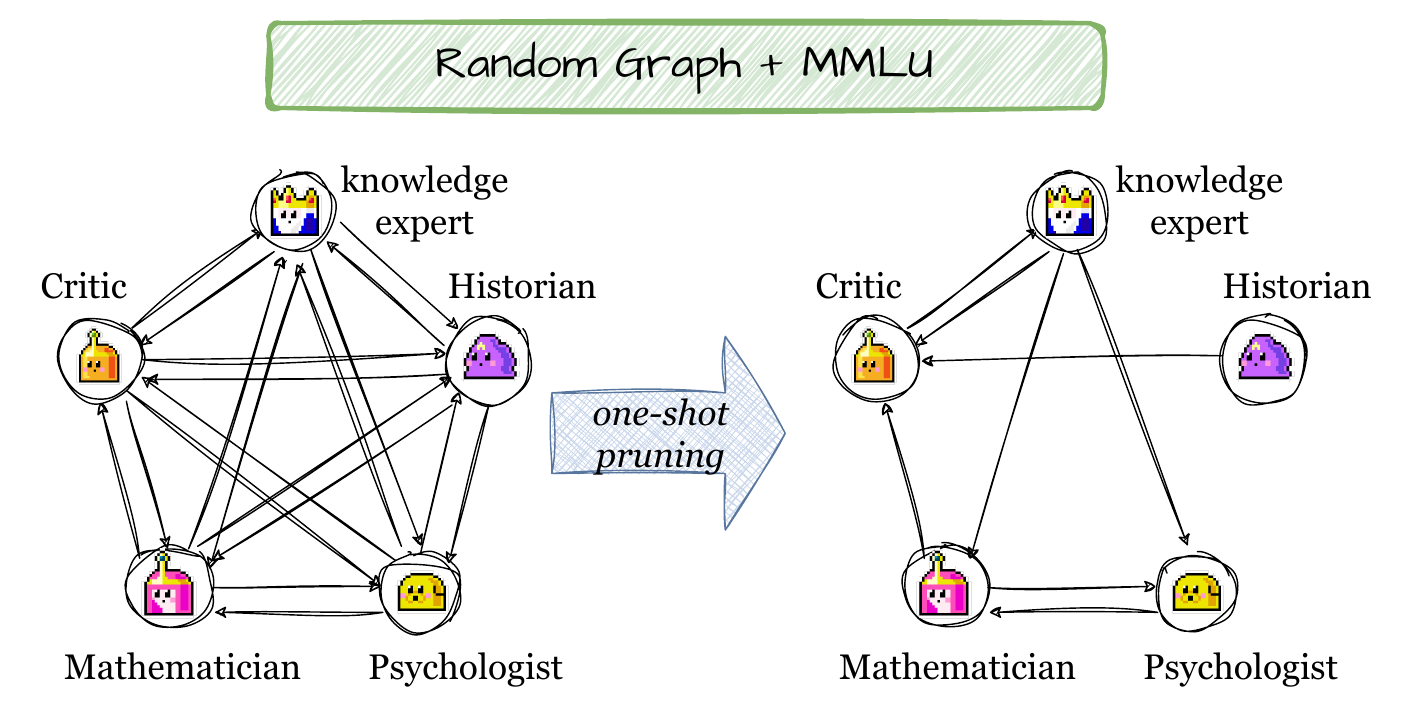}
\caption{The visualization of how \ourmethod prunes a five-agent random-graph-like framework when tested on MMLU dataset.}
\label{fig:case_random_mmlu}
\end{figure*}

\paragraph{Random Graph + MMLU} 
\Cref{fig:case_random_mmlu} illustrates how \ourmethod performs one-shot pruning, given five predefined agent roles along with a predefined \textbf{random graph} structure on \textbf{MMLU} dataset. The pruned graph structure is similarly aligned with that in \Cref{fig:case_complete_mmlu}, with the Critic having many incoming edges and the Knowledge Expert having many outgoing edges. Notably, the absence of outgoing edges for the psychologist in \Cref{fig:case_random_mmlu} may suggest that this role has limited utility within the overall system.

\begin{figure*}[!h]
\centering
\includegraphics[width=0.9\linewidth]{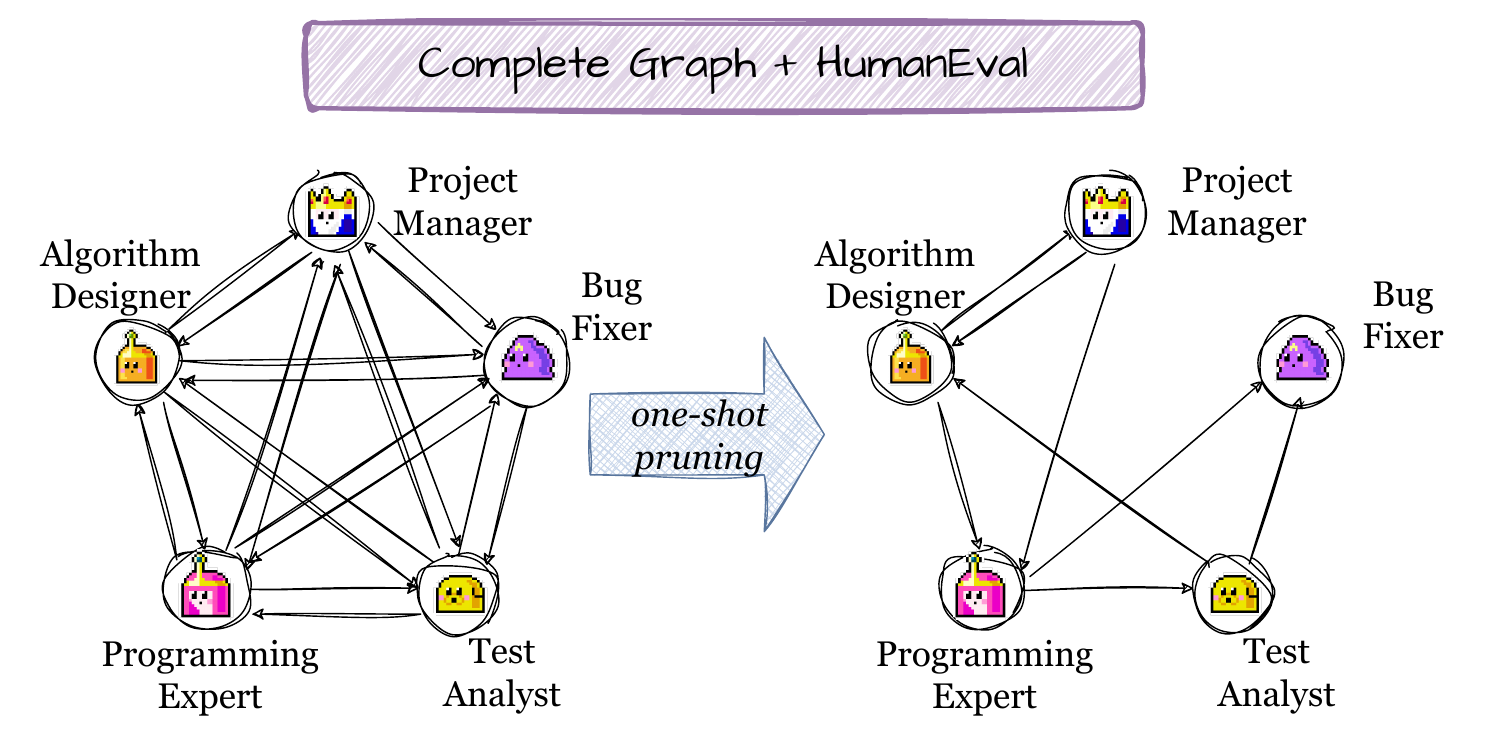}
\caption{The visualization of how \ourmethod prunes a five-agent complete-graph-like framework when tested on the HumanEval dataset.}
\label{fig:case_complete_human}
\end{figure*}

\paragraph{Random Graph + HumanEval} 
\Cref{fig:case_complete_human} demonstrates how \ourmethod implements one-shot pruning based on five predefined agent roles: project manager, algorithm designer, bug fixer, test analyst, and programming expert, utilizing a predefined \textbf{complete graph} structure on the \textbf{HumanEval} dataset. Notably, the outer edges of the graph are retained, aligning with the standard workflow for code completion, which is consistent with the designs of multi-agent code generation frameworks like MetaGPT~\citep{meta-gpt}. The Bug Fixer has no outgoing edges, as it represents the final step in the code completion process. This effectively highlights \ourmethod's capability for autonomous optimization of multi-agent collaborative topologies.

\begin{figure*}[!h]
\centering
\includegraphics[width=0.9\linewidth]{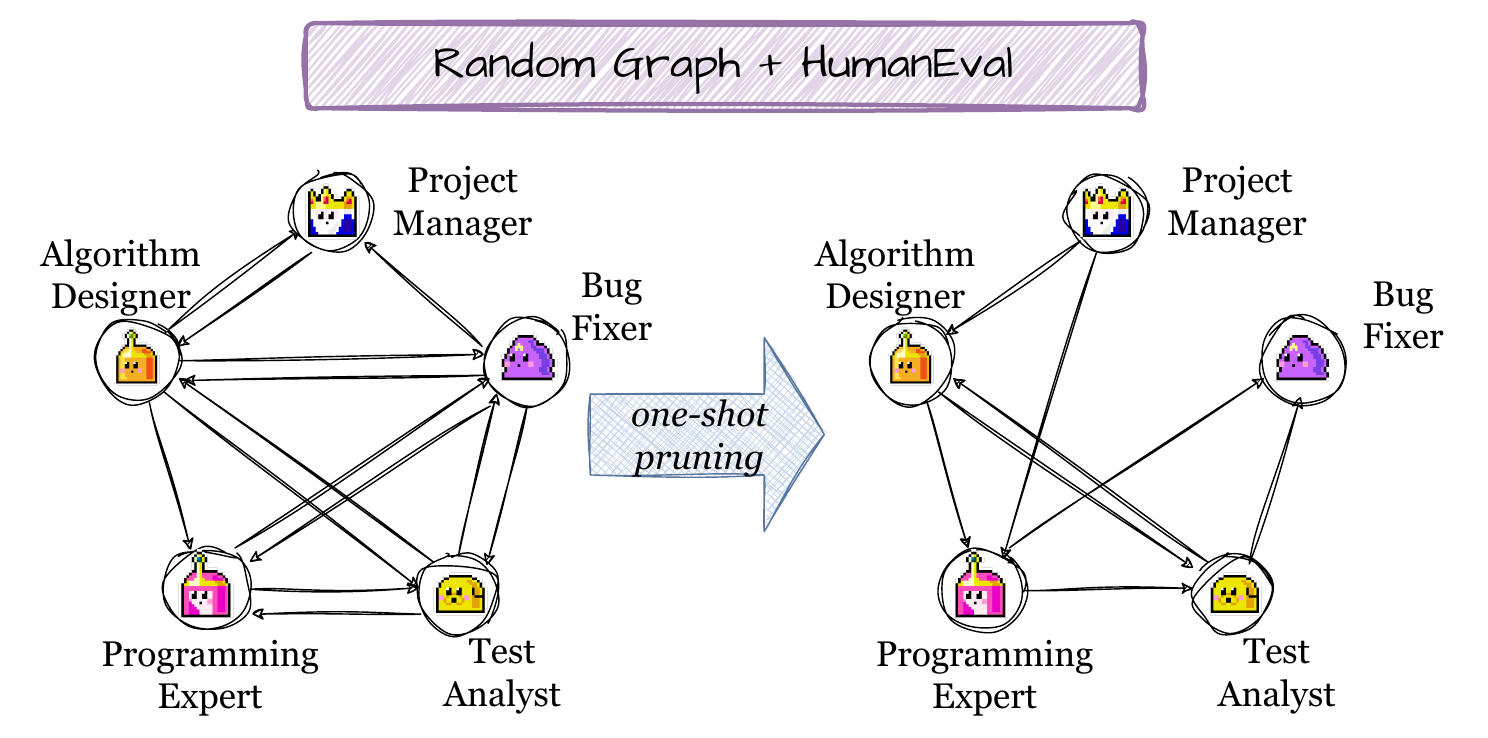}
\caption{The visualization of how \ourmethod prunes a five-agent random-graph-like framework when tested on the HumanEval dataset.}
\label{fig:case_random_human}
\end{figure*}

\paragraph{Complete Graph + HumanEval} 
\Cref{fig:case_complete_human} illustrates how \ourmethod executes one-shot pruning based on four predefined agent roles: math solver, math analyst, programming expert, and inspector, utilizing a predefined \textbf{complete graph} structure on the \textbf{GSM8K} dataset. In this scenario, we designated two agents as math solvers. Interestingly, the pruned graph structure reveals explicit differentiation in roles for the two math solvers: one is responsible for preliminary solving, while the other is tasked with final solving. The final solver gathers information from all other nodes and has no outgoing edges.

\begin{figure*}[!h]
\centering
\includegraphics[width=0.9\linewidth]{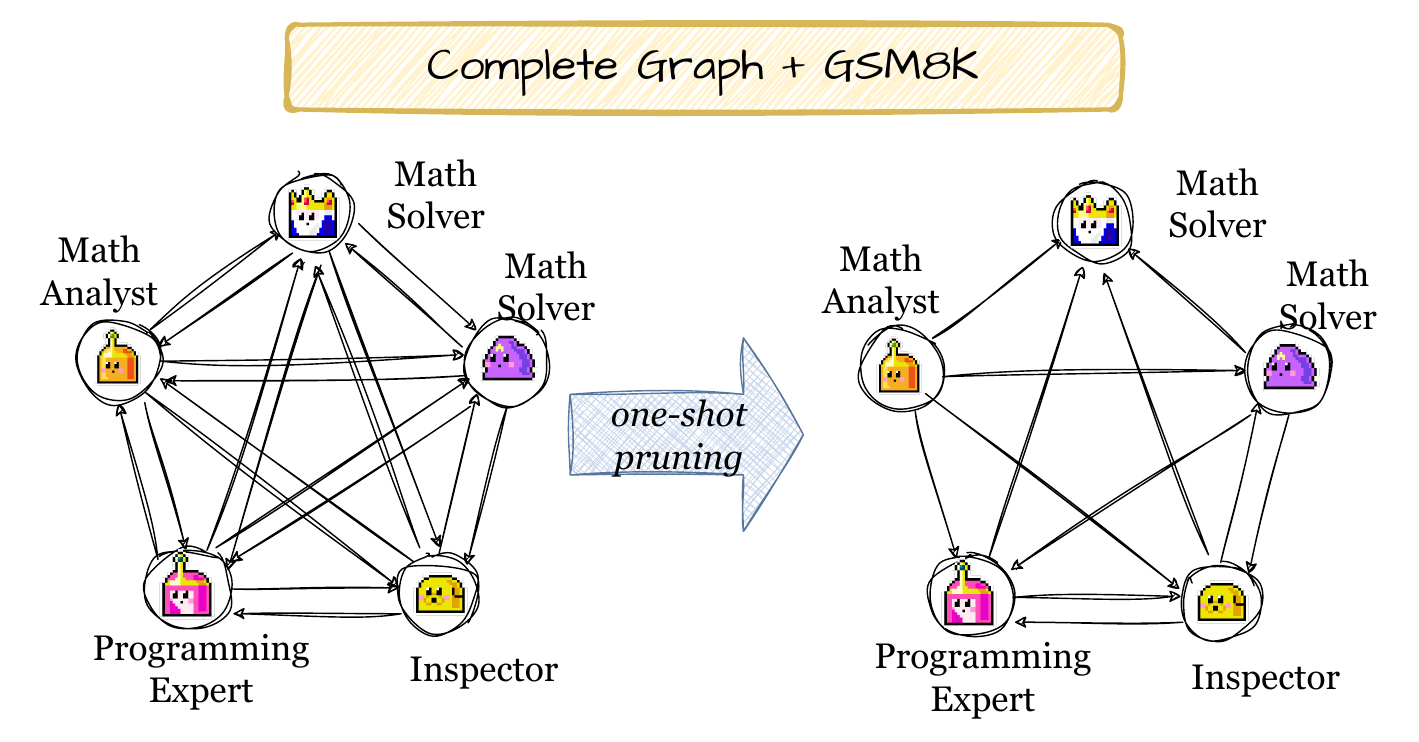}
\caption{The visualization of how \ourmethod prunes a five-agent complete-graph-like framework when tested on the GSM8K dataset.}
\label{fig:case_complete_gsm}
\end{figure*}

\paragraph{Complete Graph + GSM8K} 
\Cref{fig:case_complete_gsm} illustrates how \ourmethod performs one-shot pruning based on four predefined agent roles and a predefined \textbf{random graph} structure on the \textbf{GSM8K} dataset. We observe a distinct differentiation in node characteristics: the agents focused on problem analysis exhibit a high out-degree with no incoming edges, while the agents responsible for problem-solving demonstrate a high in-degree and a low out-degree.

\begin{figure*}[!h]
\centering
\includegraphics[width=0.9\linewidth]{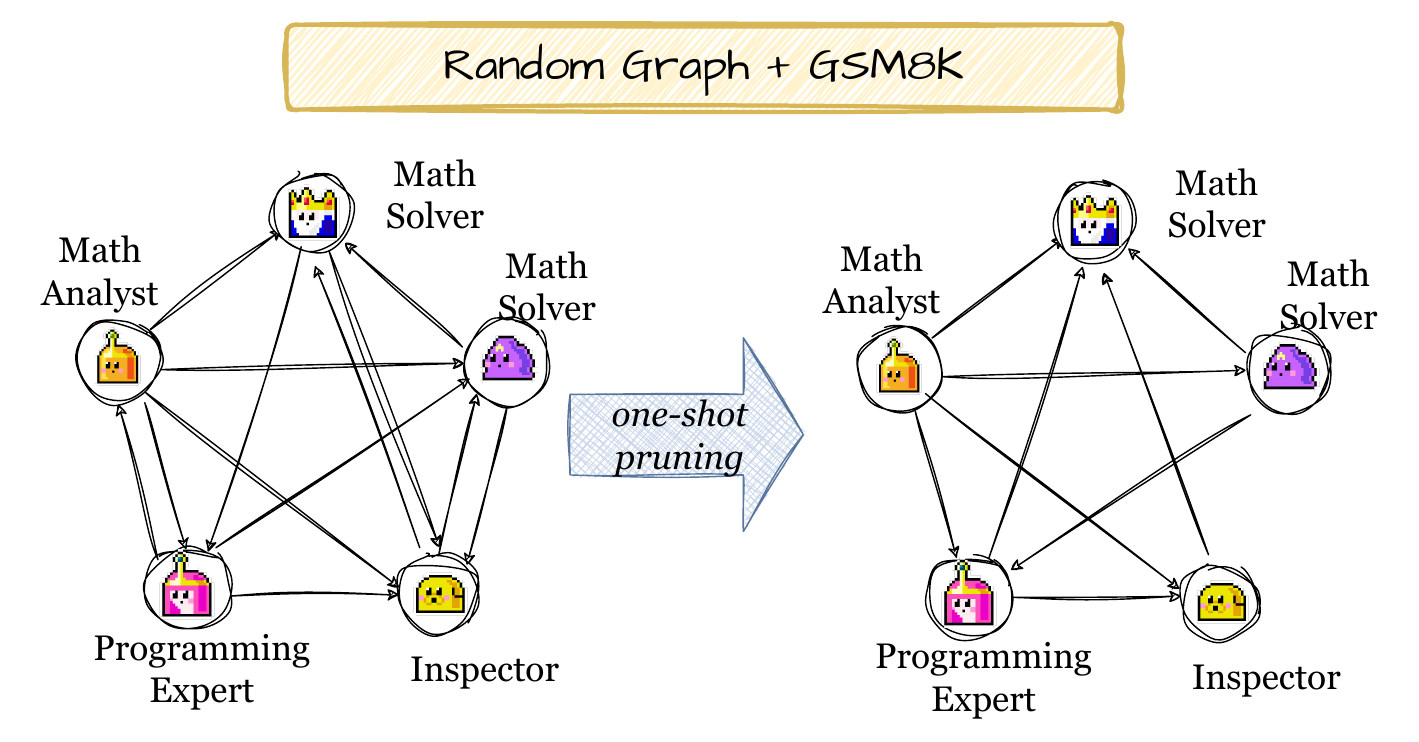}
\caption{The visualization of how \ourmethod prunes a five-agent random-graph-like framework when tested on the GSM8K dataset.}
\label{fig:case_random_gsm}
\end{figure*}

\subsection{Temporal Topology Visualization}

\begin{figure*}[!h]
\centering
\includegraphics[width=1\linewidth]{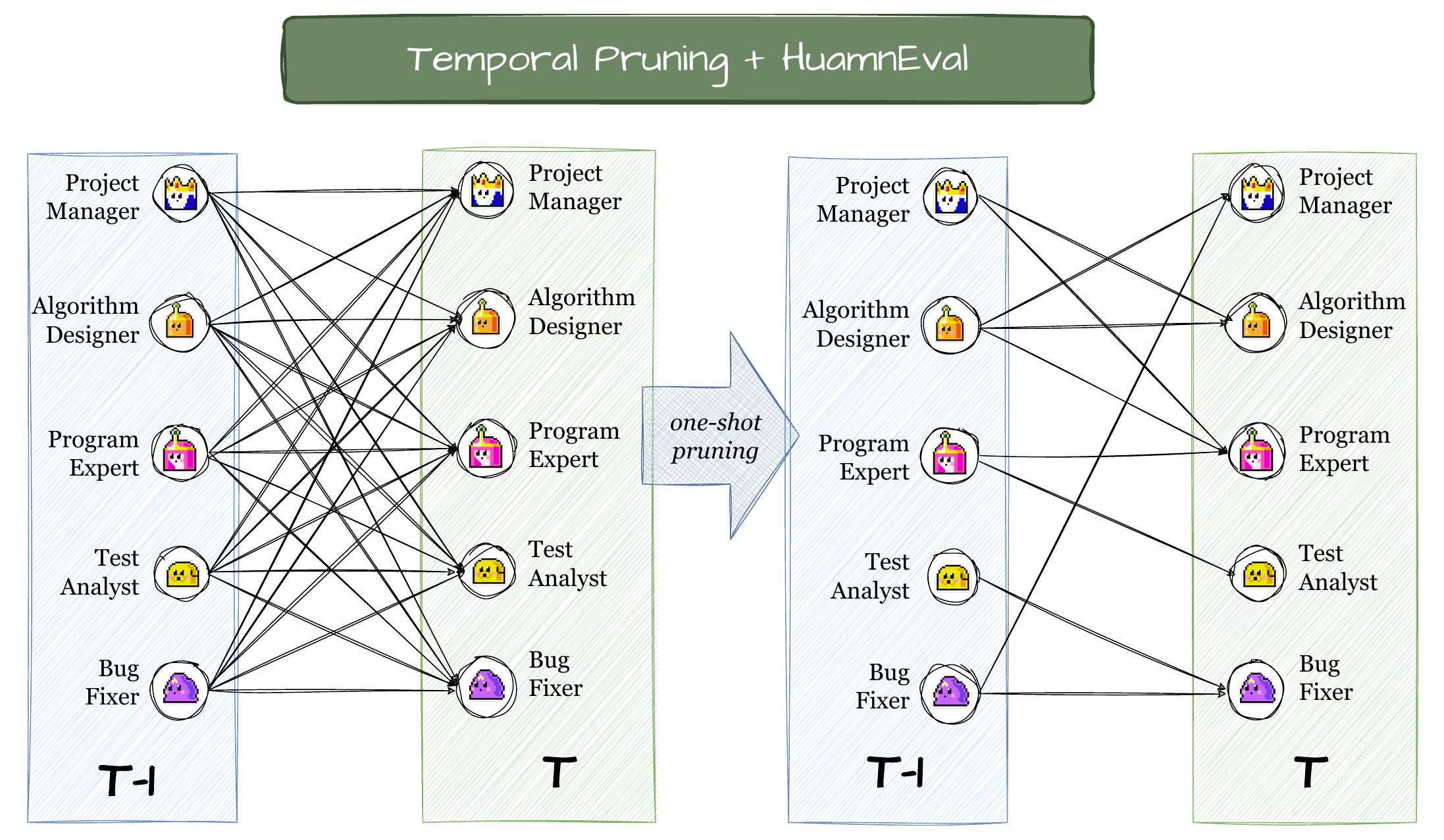}
\caption{The visualization of how \ourmethod temporally prunes a five-agent LLM-Debate-like framework when tested on the HumanEval dataset.}
\label{fig:case_temporal_human}
\end{figure*}

\begin{figure*}[!h]
\centering
\includegraphics[width=1\linewidth]{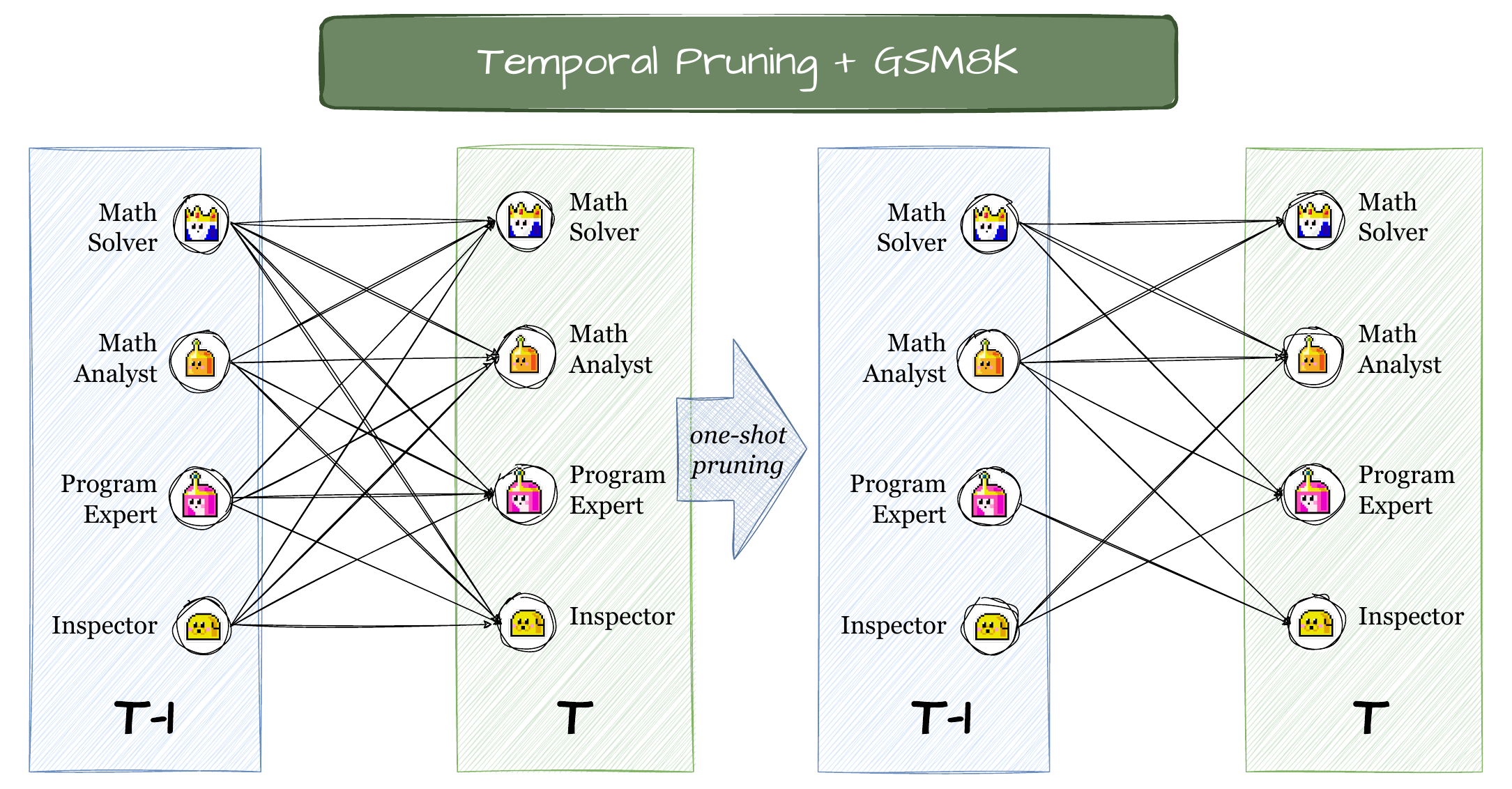}
\caption{The visualization of how \ourmethod temporally prunes a four-agent LLM-Debate-like framework when tested on the GSM8K dataset.}
\label{fig:case_temporal_gsm8k}
\end{figure*}

\Cref{fig:case_temporal_gsm8k,fig:case_temporal_human} demonstrate how \ourmethod operates on the predefined temporal communication structure and the resulting structure after one-shot pruning.

\subsection{Spatial Connectivity Optimization}
\begin{figure*}[!htbp]
\centering
\includegraphics[width=\linewidth]{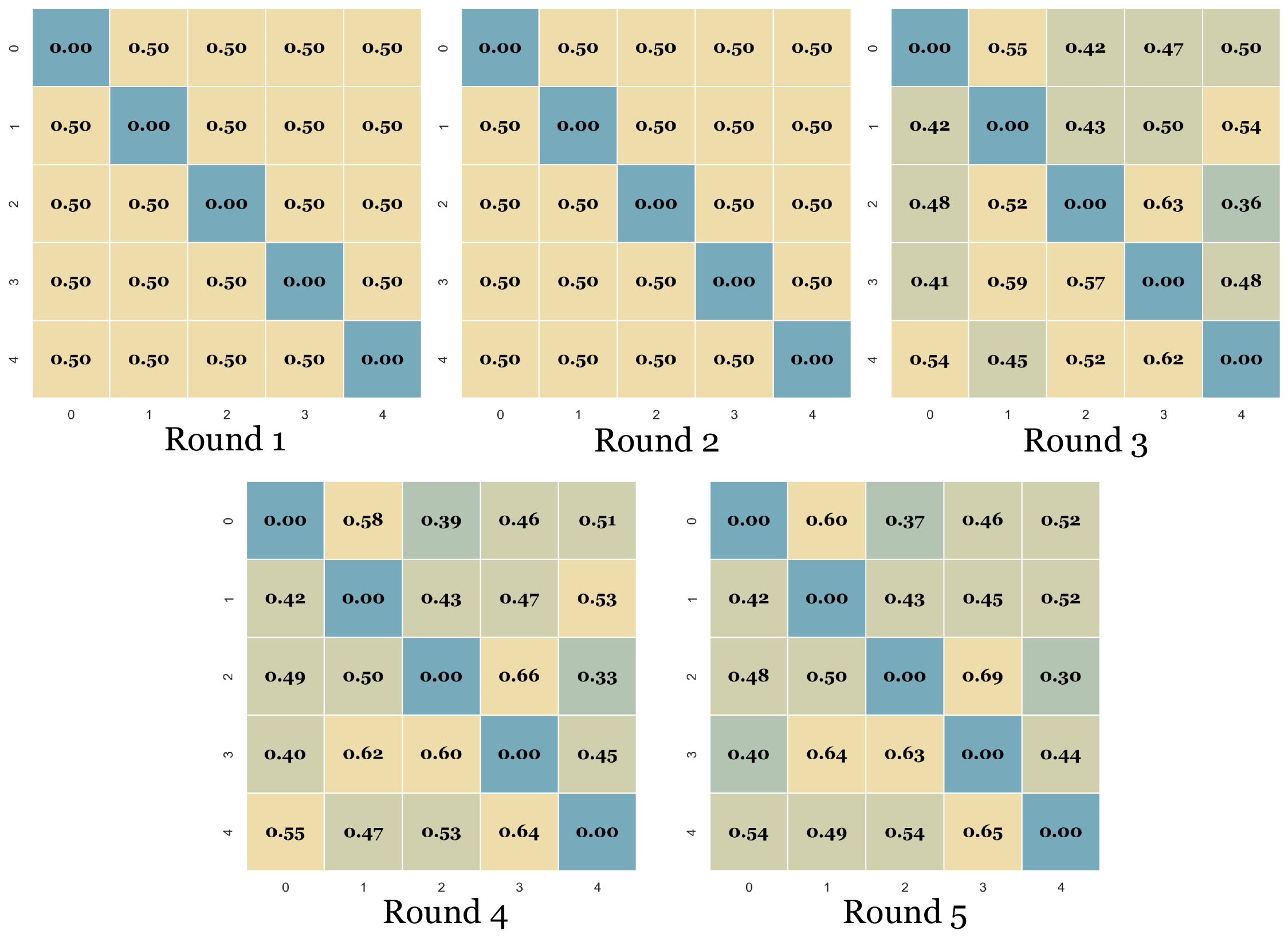}
\vspace{-1.5em}
\caption{The demonstration of how spatial connectivity evolves and optimizes over time on MMLU.}
\label{fig:adj_mmlu}
\vspace{-1.em}
\end{figure*}

\begin{figure*}[!htbp]
\centering
\includegraphics[width=\linewidth]{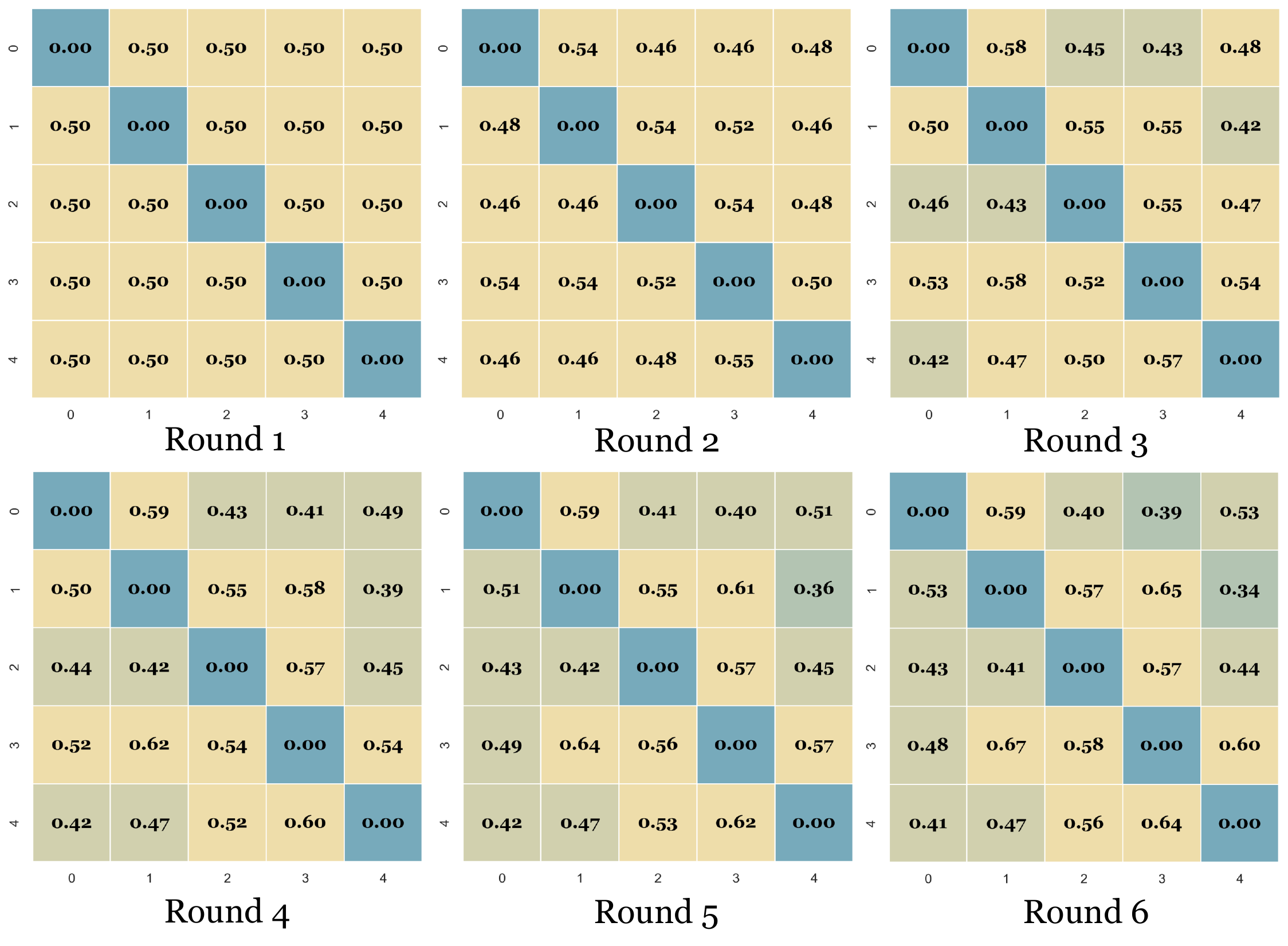}
\vspace{-1.5em}
\caption{The demonstration of how spatial connectivity evolves and optimizes over time on HumanEval.}
\label{fig:adj_human}
\vspace{-1.em}
\end{figure*}

\begin{figure*}[!htbp]
\centering
\includegraphics[width=\linewidth]{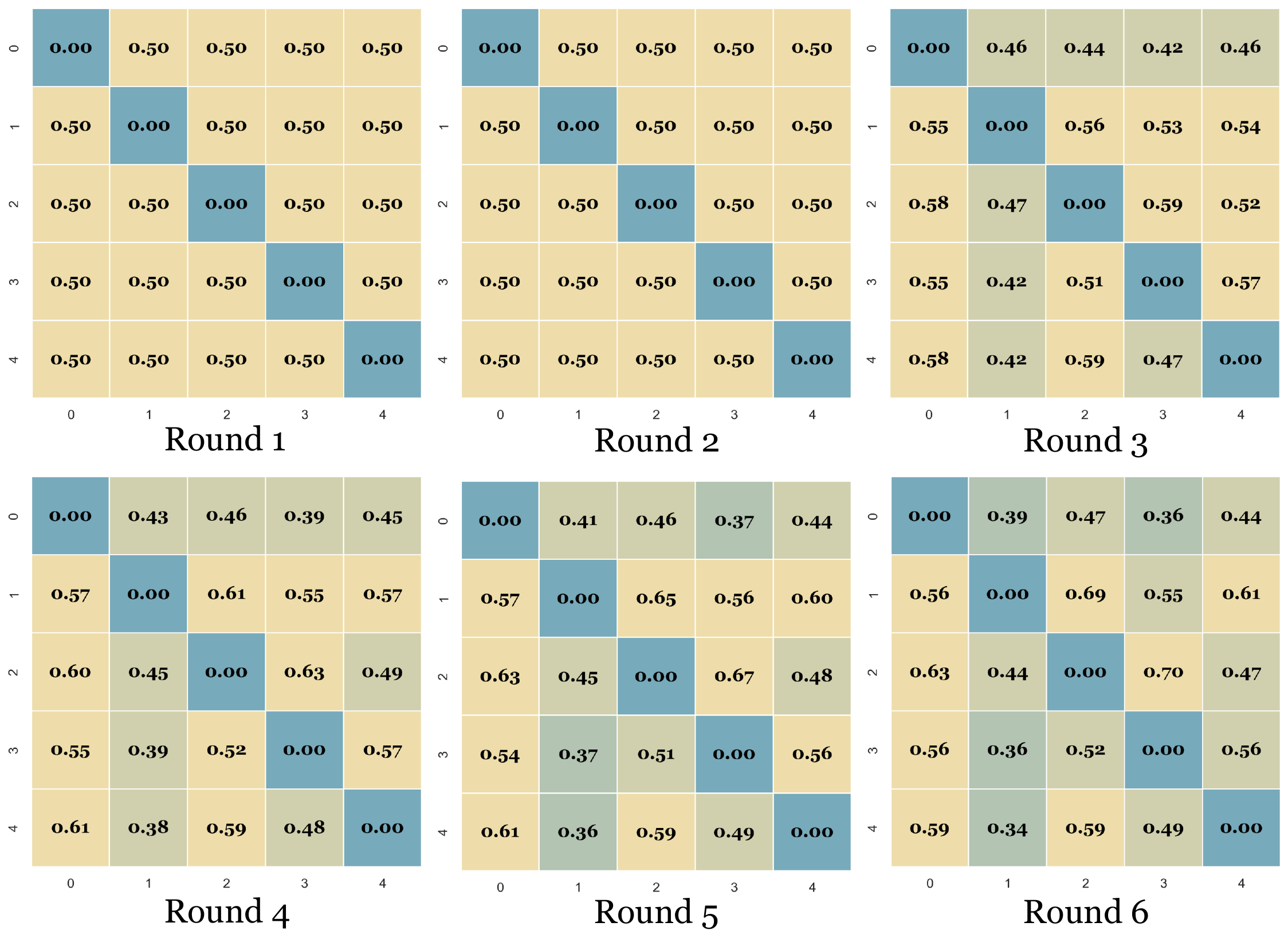}
\vspace{-1.5em}
\caption{The demonstration of how spatial connectivity evolves and optimizes over time on GSM8K dataset.}
\label{fig:adj_gsm8k}
\vspace{-1.em}
\end{figure*}

\Cref{fig:adj_mmlu,fig:adj_gsm8k,fig:adj_human} illustrate the optimization trajectory of spatial connectivity before applying one-shot pruning. As time progresses, the initially identical graph masks begin to show increasing differentiation, with their magnitudes serving as a crucial reference for subsequent pruning decisions.

\end{document}